\documentclass[12pt]{article}
\usepackage{amsmath}
\usepackage{amssymb}
\usepackage{amscd}
\usepackage{pstricks}
\usepackage{pst-all}
\usepackage[all]{xy}
    \SelectTips{cm}{}
\usepackage[mathscr]{euscript}
\usepackage{graphicx}
\usepackage{booktabs}
\usepackage{theorem}
\theoremstyle{plain}
\newtheorem{thm}{Theorem}[section]
\newtheorem{prp}[thm]{Proposition}
\newtheorem{dfn}[thm]{Definition}

\newcommand{\qed}{\hbox{\rule[-2pt]{3pt}{6pt}}}

\usepackage{cite}

\makeatletter
 \@addtoreset{equation}{section}
 
 \makeatother

\begin{document}

\title{Topos Quantum Theory Reduced by  Context-Selection Functors}
\author{Kunji Nakayama\footnote{e-mail: nakayama@law.ryukoku.ac.jp}
\\
Faculty of Law\\
Ryukoku University\\
Fushimi-ku, Kyoto 612-8577}

\maketitle

\def\Rmath{{\mathbb{R}}}

\def\Set{{\bf Set}}
\def\Sets{{\bf Sets}}
\def\Sh{{\rm Sh}}
\def\op{{\rm op}}

\def\Ocal{{\cal O}}
\def\Ocalop{{\cal O}^{\rm op}}
\def\OcalBorelR{\Ocal \times \mathfrak{B}_{\mathbb{R}}}
\def\Obj{{\rm Obj}}
\def\Mor{{\rm Mor}}

\def\Ahat{\hat{A}}
\def\Bhat{\hat{B}}
\def\Ehat{\hat{E}}
\def\Hhat{\hat{H}}
\def\Rhat{\hat{R}}
\def\hhat{\hat{h}}
\def\khat{\hat{k}}
\def\lhat{\hat{l}}
\def\gammahat{\hat{\gamma}}

\def\Aboldhat{\widehat {\bf A}}
\def\Eboldhat{\widehat {\bf E}}
\def\Cboldhat{\widehat {\bf C}}
\def\Pboldhat{\widehat {\bf P}}
\def\Vboldhat{\widehat {\bf V}}
\def\Vcalhat{\widehat {\cal V}}

\def\Aboldop{{\bf A^{{\rm op}}}}
\def\Cboldop{{\bf C^{{\rm op}}}}
\def\Eboldop{{\bf E^{{\rm op}}}}
\def\Vboldop{{\bf V^{{\rm op}}}}
\def\Vcalop{{\cal V^{{\rm op}}}}

\def\Aboldop{{\bf A}^{\rm op}}

\def\Vboldupsilon{{\bf V}_{\uptilde}}
\def\Vboldupsilonop{{\bf V}_{\uptilde}^{\rm op}}
\def\Vboldupsilonhat{\hat{{\bf V}_{\uptilde}}}

\def\CboldOcal{{\bf C}_{{\cal O}}}
\def\CboldOcalop{{\bf C}_{{\cal O}}^{\rm op}}
\def\CboldOcalhat{\widehat{{\bf C}_{{\cal O}}}}

\def\alphabold{\boldsymbol{\alpha}}
\def\betabold{\boldsymbol{\beta}}

\def\ybold{{\bf y}}
\def\Abold{{\bf A}}
\def\abold{\boldsymbol{ a}}

\def\Bbold{{\bf B}}
\def\Cbold{{\bf C}}
\def\Gbold{{\bf G}}
\def\Ebold{{\bf E}}
\def\Fbold{{\bf F}}
\def\Mbold{{\bf M}}
\def\Nbold{{\bf N}}
\def\Kbold{{\bf K}}
\def\Hbold{{\bf H}}
\def\Lbold{{\bf L}}
\def\Pbold{{\bf P}}
\def\Qbold{{\bf Q}}
\def\Vbold{{\bf V}}
\def\Sbold{{\bf S}}
\def\Tbold{{\bf T}}
\def\Ibold{{\bf I}}
\def\Jbold{{\bf J}}
\def\jbold{{\bf j}}
\def\Omegabold{\boldsymbol{\Omega}}
\def\omegabold{\boldsymbol{\omega}}
\def\Sigmabold{\boldsymbol{\Sigma}}
\def\zetabold{\boldsymbol{\zeta}}
\def\Kcal{{\cal K}}

\def\Pcal{{\cal P}}
\def\Qcal{{\cal Q}}
\def\betatilde{\tilde{\beta}}

\def\Subsets{{\rm Subsets}}

\def\Hom{{\rm Hom}}
\def\Sub{{\rm Sub}}
\def\Hyp{{\rm Hyp}}
\def\dom{{\rm dom}}
\def\cod{{\rm cod}}
\def\Hyp{{\rm Hyp}}

\def\Flt{{\cal F}}
\def\Dcal{{\mathfrak D}}
\def\Ecal{{\mathfrak E}}
\def\Fcal{{\mathfrak F}}
\def\Lcal{{\mathfrak L}}
\def\Ocal{{\cal O}}
\def\Scal{{\cal S}}

\def\Amath{{\mathfrak A}}
\def\Bmath{{\mathfrak B}}
\def\Cmath{{\mathfrak C}}
\def\Emath{{\mathfrak E}}

\def\Scalbar{{\bar{\cal S}}}
\def\Hcal{{\cal H}}
\def\Ccal{{\cal C}}
\def\Gcal{{\cal G}}
\def\Rcal{{\cal R}}

\def\Tcal{{\cal T}}
\def\Vcal{{\cal V}}
\def\tmath{{\mathfrak{t}}}
\def\Dmath{{\mathfrak D}}
\def\Fmath{{\mathfrak F}}
\def\Vmath{{\mathfrak V}}

\def\CcalR{{{\cal C}_R}}
\def\CcalRe{{{\cal C}_R^{e \downarrow}}}
\def\Ocaltilde{{\cal O}/_{\sim}}

\def\ahat{\hat{a}}
\def\Fhat{\hat{F}}
\def\Phat{\hat{P}}
\def\Hhat{\hat{H}}
\def\Ohat{\hat{O}}
\def\Ihat{\hat{I}}
\def\alphahat{\hat{\alpha}}

\def\Ghat{\hat{G}}

\def\Com{{\rm Com}}
\def\ES{{\rm ES}}
\def\op{{\rm op}}
\def\Onebold{{\bf 1}}
\def\Mor{{\rm Mor}}
\def\Obj{{\rm Obj}}
\def\Bcal{{\cal B}}
\def\End{{\rm End}}

\def\deltabold{\boldsymbol{\delta}}
\def\deltaboldj{\boldsymbol{\delta}_{j}}

\def\Cmathbb{\mathbb{C}}
\def\Pmathbb{\mathbb{P}}
\def\Rmathbb{\mathbb{R}}
\def\Smathbb{\mathbb{S}}
\def\Tmathbb{\mathbb{T}}
\def\Tmathbbphi{\mathbb{T}^{| \varphi \rangle}}
\def\Tmathbbrhor{\mathbb{T}^{\rho, \, r}}

\def\varphivec{| \varphi \rangle}
\def\varphipro{| \varphi \rangle \langle \varphi |}

\def\taurhor{\tau^{\rho, \, r}}
\def\tauphi{\tau^{| \varphi \rangle}}

\def\subseteqVbold{\subseteq_{\Vbold}}

\def\true{{\rm true}}
\def\onepoint{\{ \, \cdot \,\}}
\def\id{{\rm id}}
\def\pt{{\rm pt}}
\def\dB{{\rm dB}}
\def\cl{{\rm cl}}
\def\Cl{{\cal C}l}

\def\alphabreve{\breve{\alpha}}
\def\uptilde{\tilde{\upsilon}}

\def\vertin{\mathop{{\rotatebox{90}{$\in$}}}}

\def\vertsim{\mathop{{\rotatebox{90}{$\sim$}}}}
\def\fortyfivesim{\mathop{{\rotatebox{-45}{$\sim$}}}}

\def\tr{{\rm tr}}
\def\qed{\hspace{\fill}$\square$\\}

\def\Omegabold{\boldsymbol{\Omega}}
\def\omegabold{\boldsymbol{\omega}}
\def\Sigmabold{\boldsymbol{\Sigma}}
\def\zetabold{\boldsymbol{\zeta}}
\def\Kcal{{\cal K}}
\def\gammabold{\boldsymbol{\gamma}}
\def\gammaboldmax{\boldsymbol{\gamma}^{\vee}}
\def\gammaboldmin{\boldsymbol{\gamma}^{\wedge}}
\def\jmathboldmax{\boldsymbol{\jmath}^{\vee}}
\def\jmathboldmin{\boldsymbol{\jmath}^{\wedge}}
\def\jmathbold{\boldsymbol{\jmath}}
\def\imathbold{\boldsymbol{\imath}}
\def\imathboldmax{\boldsymbol{\imath}^{\vee}}
\def\imathboldmin{\boldsymbol{\imath}^{\wedge}}

\def\Tpre{{\rm Sub}_{{\rm filt}} (\Pmathbb_{\cl} \Sigma)}

\def\Ucalflat{\mathcal{U}^{\flat}}

\def\Doering{{D\"{o}ring}}
\def\prob{{\rm p}}
\def\Prob{{\rm Prob}}

\begin{abstract}

In this paper, we deal with quantum theories on presheaves and sheaves
on context categories consisting of commutative von Neumann algebras 
of bounded operators on a Hilbert space,
from two viewpoints.
One is to reduce presheaf-based topos quantum theory via sheafification,
and the other is to import quantum probabilities to the reduced sheaf quantum theory.
The first is done by means of a functor that selects some expedient contexts.
It defines a Grothendieck topology on the category consisting of all contexts,
hence, induces a sheaf topos 
on which we construct a downsized quantum theory.
Also, we show that the sheaf quantum theory can be replaced by 
an equivalent, more manageable presheaf quantum theory.
Quantum probabilities are imported by means of a Grothendieck topology that is defined on a category consisting of probabilities and enables to regard them as intuitionistic truth-values.
From these topologies,
we construct another Grothendieck topology that is defined on the product of the context category
and the probability category and reflects the selection of contexts and 
the identification of probabilities with truth-values.
We construct a quantum theory equipped with quantum probabilities as truth-values 
on the sheaf topos induced by the Grothendieck topology.

\end{abstract}

\newpage

\section{Introduction}

Topos quantum theory is a truth-value oriented approach similar to classical physics
\cite{I97,IB98,BI99,HBI00,BI02,DI08a,DI08b,DI08c,DI08d,HLS09,F10,HLS11,HLSW11,DI11,W13,F13a,F13b}. 
Any classical theory could be regarded as an inference system that gives 
a 2-valued truth-value to every physical proposition concerning a value of a physical quantity.
Similarly, topos quantum theory assigns an intuitionisic truth-value
to every such physical proposition concerning a quantum system,
instead of predicting a probability with which experiments realize the proposition.
Consequently, topos approach makes it possible to
establish quantum theory without the notion of measurement,
hence, without any dichotomy between an observer and an observed system.
Because of this, 
topos approach is expected to provide a consistent framework promising for quantum gravity and quantum cosmology.
In order for the approach to be valid, however,
it needs to be applicable to concrete quantum systems.
Furthermore,  to challenge the tough problems, formulation of quantum gravity theory and quantum cosmology,
it may be as well to accumulate experience in investigating various quantum systems
in a topos theoretic framework.
Nonetheless, topos approach is still in a stage of construction of general theory
and there are few application studies.
One possible reason of this would be in the formulation itself.

To see the reason,
let us observe the essential feature of topos quantum theory
taking the theory developed by D\"{o}ring and Isham \cite{DI08a,DI08b,DI08c,DI08d,DI11}.
They used the topos of presheaves on the category of contexts,
i.e., the category $\Vbold$ of commutative von Neumann algebras of bounded operators on a Hilbert space $\Hcal$
with morphisms defined by inclusion relation.
They found that the spectral presheaf $\Sigma$, which assigns to each context $V \in \Vbold$ the Gel'fand space $\Sigma(V)$ on $V$,
plays a role similar to state space of classical physics.
Every proposition on a classical system can be represented by a collection of  extentions,
namely, the subset of phase space each point of which makes the proposition true.
On the other hand, in the topos quantum theory,
every physical proposition on a quantum system is represented by a clopen sub-presheaf of $\Sigma$,
though $\Sigma$ has no points
because of the Kochen-Specker no-go theorem\cite{IB98,BI99,HBI00,BI02,KS67}.
If we are given a vector state $\varphivec$ or a density matrix $\rho$ of the quantum system,
we can define a corresponding truth presheaf, which gives propositions that the system makes true.
Then, we can assign a truth-value to any proposition by the standard method known in topos theory.
The truth-values are taken on the Heyting algebra $\Gamma \Omega$ consisting of global elements of the sub-object classifier $\Omega$,
where, for each contest $V$, $\Omega(V)$ consists of all sieves on $V$.

As it turns out,
in the presheaf quantum theory of D\"{o}ring and Isham, 
all of the contexts are evenly treated at the same time.
Because of this, when applying the theory to concrete situations,
we have to deal with vast number of truth-values that relate mutually to constitute a Heyting algebra. 
For example, when we deal with a many-particle system,
we have to treat bounded operators on a tensor-product of Hilbert spaces.
Then, the more the number of particles increases, the more the number of such operators,
the more the number of contexts,
hence, the huger the space  $\Gamma \Omega$ of truth-values.
Suppose we are interested, for instance, only in the spin components in specific directions.
Then, most of the contexts do not relate to the corresponding operators,
hence most components of truth-values would be inessential
since they are evaluated on the contexts unrelated to such operators. 
In general, whole spaces of contexts and truth-values are needed to construct a general theory of a quantum system.
However,
they may be too big when our interest is confined only to a specific part of properties of the system.
Therefore, from the practical viewpoint, 
it would be meaningful to downsize the spaces so as to make a theory more manageable.
This is the first issue the present paper addresses.

As previously noted,
topos approach  makes a self-contained theory holding without the notion of measurements.
As will be shown in this paper,
this is the case also for the downsized topos quantum theory.
In order for topos approach to be valid, however,
it has to give consistent results with ordinary quantum physics.
In particular, it has to be able to interpret quantum probabilities predicted by ordinary theory.
This point is already answered affirmatively by D\"{o}ring and Isham \cite{D09,DI12}
for the presheaf quantum theory.
D\"{o}ring \cite{D09} defined a measure on the spectral presheaf.
He showed that any qunatum state expressed by a density matrix $\rho$
induces a measure $\mu^{\rho}$ on $\Sigma$,
by which quantum probabilities predicted by ordinary quantum theory
can be reproduced.
Furthermore, D\"{o}ring and Isham \cite{DI12} showed that
topos theory can treat quantum probabilities as intuitionistic truth-values,
though the probabilities should be regarded 
as representing not relative frequency but propensity or potentiality.
Consequently, the presheaf-based quantum theory of D\"{o}ring and Isham
\cite{DI08a,DI08b,DI08c,DI08d,DI11} seems to have more information concerning a quantum system,
i.e., truth-values and probabilities,
than ordinary quantum theory.
In order for the downsized topos quantum theory to work well,
it has to inherit this favorable charateristic from the presheaf quantum theory.
To show this is the second issue of this paper.

As mentioned above, this paper addresses two main issues;
One is to downsize topos quantum theory
and the other is to import quantum probabilities to the reduced theory.
The key idea for the first is to select some contexts from the context category so as to fit one's purpose.
This is realized by means of an endofunctor on $\Vbold$,
which we call a context-selection functor, or shortly, a context selector.
Such a functor induces sheaf topos, 
on which sheaf-based quantum theory consisting of reduced theoretical ingredients can be constructed. 
This idea is a generalization of the sheaf-based quantum theory of Nakayama \cite{N13,N14},
where the original presheaf-based topos quantum theory by  D\"{o}ring and Isham is largely reduced
by selecting specific contexts by means of a quantization map.
So, we can apply whole of the formulation method developed by Nakayama \cite{N13,N14}.
The advantage of this formulation is in that the reduced theory is given as a sub-theory
of the original presheaf theory.
Consequently, the relations between the theoretical ingredients, say, 
truth-values assigned by each theory to the same physical proposition, are easily made clear.
However, we need to note limitations of the representational power of the reduced theory, 
because it is obtained as a coarse graining of the presheaf theory.
On the other hand, in order to bring quantum probabilities into the reduced topos quantum theory, 
we generalize the methods developed by
D\"{o}ring and Isham \cite{D09,DI12} to the sheaf-based regime.
Accordingly, this paper can be regarded as a sequel that converges the papers by D\"{o}ring and Isham \cite{D09,DI12} and Nakayama \cite{N13,N14}.

This paper is organized as follows. 
In section \ref{sec:review of Nakayama 2014}, we construct a reduced topos quantum theory.
This is, in fact, just a briefly review of
sheaf-based topos quantum theory of Nakayama \cite{N13,N14}.
The functor $\flat$, however, has a largely extended meaning as an operator-selection functor.
We show that physical propositions on values of physical quantities contained 
in a context selected by $\flat$ are faithfully represented 
by clopen sub-sheaves of the spectral sheaf. 
Accordingly, 
though the representational power of the reduced theory is less than the presheaf theory,
we could safely say that the former works well at least for such physical quantities.
Section \ref{sec:probability measure} can be regarded as an extension of the measure theory of D\"{o}ring \cite{D09}.
We define measures on the spectral sheaf, i.e.,
the sheafification of the spectral presheaf, and show that they can reproduce correct quantum probabilities
for physical propositions on the selected operators.
Since the measure is defined as a notion external to the sheaf topos,
we define morphisms representing measures in the topos,
and we show some aspects of truth objects and truth-value valuation.
Sections \ref{sec:topology for quantum probability} and \ref{sec:ibold-sheaf topos} can be regarded as 
an extension of the topos theoretic treatment of classical and quantum probabilities developed by D\"{o}ring and Isham \cite{DI12}.
In section 4, we reconsider the method by which probabilities can be seen as intuitionistic truth-values.
We construct a sheaf topos from a presheaf topos defined on the space of probabiliries $[0,1]$
via defining a suitable Grothendieck topology on $[0,1]$.
In section \ref{sec:ibold-sheaf topos},  
we join the topology induced by $\flat$ and that of probabilities to define a Grothendieck topology
on the category defined as a product of the context space and the probability space.
This topology induces a sheaf topos on which we can define a quantum theory provided with probabilities as truth-values.
This is an extention of the theory of D\"{o}ring and Isham \cite{DI12}
to the regime with some contexts selected.
In section \ref{presheaf-based reduced theory},
we return to the topos quantum theory given in section \ref{sec:review of Nakayama 2014}.
Though we can easily relate the sheaf-based theory to the presheaf-based one of  D\"{o}ring and Isham,
the former is still somewhat redundant,
since sheaves are defined on all of the context category,
whereas only the values on $\flat(\Vbold)$ are needed to determine them.
We give a presheaf quantum theory on $\flat(\Vbold)$ that is equivalent to the sheaf theory.


\section{Reduced quantum theory on sheaves induced by a context-selection functor}
\label{sec:review of Nakayama 2014}
\subsection{The $j$-sheaf topos quantum theory}

In this subsection, we give a topos quantum theory on sheaves induced by a selection of contexts.
This is just a brief summary of Nakayama \cite{N14}.
However,  the meaning of $\flat$ is largely generalized.
We adopt the notation of Nakayama \cite{N14} here and hereafter.
That is, the Hilbert space on which a physical system and observables are represented
is denoted by $\Hcal$.
The category of all contexts, each of which is a commutative von Neuman algebra of
bounded operators on $\Hcal$, is denoted by $\Vbold$,
where morphisms are defined by inclusion, i.e.,
$V' \hookrightarrow V \in \Mor(\Vbold)$ if and only if $V$ includes $V'$.
(We denote the inclusion relation by $V' \subseteqVbold V$.)
The quantum theory of D\"{o}ring and Isham is constructed on  the topos $\Vboldhat \equiv \Set^{\Vbold^{\op}}$ of presheaves on $\Vbold$.

\begin{dfn}
An endofunctor $\flat:\Vbold \to \Vbold$ is said to be a context-selection functor,
or briefly, a context selector,
if it satisfies, 
\begin{description}
\item[(i)] for all $V \in \Vboldhat$,
\begin{equation}
\flat(V) \subseteqVbold V,
\label{eq:flat1}
\end{equation}
\item[(ii)] and is an idempotent:
\begin{equation}
\flat \flat = \flat.
\label{eq:flat2}
\end{equation}
\end{description}
\end{dfn}
By means of $\flat$,
we effectively select such contexts as are fixpoints of $\flat$, namely $V \in \Vbold$ such that $\flat(V)= V$,
or equivalently, $V \in \flat(\Vbold)$.
\begin{dfn}
\label{dfn:flat-selected context}
Any context $V \in \Vbold$ is said to be $\flat$-selected if $V \in \flat(\Vbold)$.
\end{dfn}
\begin{dfn}
\label{dfn:flat-selected operator}
Any bounded operator $\Ahat$ on $\Hcal$ is said to be $\flat$-selected if
it is contained in a $\flat$-selected context.
\end{dfn}
Regarding definition \ref{dfn:flat-selected operator}, 
we will also use expressions such as `$\flat$-selected observables,' `$\flat$-selected physical quantities,' and so on,
if the observables are expressed by $\flat$-selected operators.

One natural way to define a context selector is 
to choose a (finite or infinite) subset $S$ of  (commutative or noncommutative) bounded operators.
For such an $S$, we can define a context selector $\flat_{S}$ by
\begin{equation}
\flat_{S}(V) :=((S \cup S^{*})\cap V)'',
\end{equation}
where $S^{*}$ is the set of Hermitian conjugates of the  operators in $S$
and $''$ is the double commutant.
For example, Nakayama \cite{N13,N14} defined $\flat_{S}$ from quantization of classical observables.
That is, $S$ was taken to be a set of images of classical observables via a quantization map assumed to be faithful.
Also, any faithful represention of a $C^{*}$-algebra on $\Hcal$ can define a $\flat_{S}$.
Moreover, we can select an arbitrary $S$ by hand for any purpose.
In the present paper, we make no assumptions concerning how to define a context-selector;
Any $\flat$ satisfying (\ref{eq:flat1}) and (\ref{eq:flat2}) is allowed. 

Any context selector $\flat$ naturally
induces a Grothendieck topology $J$ on $\Vbold$,
which is defined by, for each $V \in \Vbold$,
\begin{equation}
J(V) := \{ \omega  \in \Omega(V)\;|\; \flat(V) \in \omega\},
\label{eq:Grothendieck}
\end{equation}
where $\Omega$ is the sub-object classifier of $\Vboldhat$.


The Lawvere-Tierney topology $\Omega \xrightarrow{j} \Omega\in \Mor(\Vboldhat)$ corresponding to the Grothendieck topology (\ref{eq:Grothendieck}) is given as, for each $V \in \Vbold$ and $\omega \in \Omega(V)$,
\begin{equation}
j_{V}(\omega):= \{ V' \subseteqVbold V \;|\; \flat(V') \in \omega \}.
\label{eq:LT}
\end{equation}
The topologies $J$ and $j$ are, furthermore, equivalent to a closure operator defined 
on the collection $\Sub(Q)$ of sub-presheaves of a presheaf $Q \in \Vboldhat$.
This is a map that assigns to each $S \in \Sub(Q)$
the closure $\bar{S}$ in $Q$ defined by
\begin{equation}
\bar{S}(V) := \{ q \in Q(V) \;|\; Q(\flat(V) \hookrightarrow V)(q) \in S(\flat(V))  \}.
\label{eq:closure}
\end{equation}

We introduce the pullback functor $\flat^{*}:\Vboldhat \to \Vboldhat$ defined by
\begin{equation}
(\flat^{*}Q)(V) := Q(\flat(V)) ,\quad
(\flat^{*}Q)(V' \hookrightarrow V) := Q(\flat(V') \hookrightarrow \flat(V)),
\end{equation}
and, for any $f \in \Mor(\Vboldhat)$,
\begin{equation}
(\flat^{*}f)_{V} := f_{\flat(V)}.
\end{equation}
Then, sheaves associated with (\ref{eq:LT}) is expressed by a natural transformation $I \xrightarrow{\zeta} \flat^{*}$ defined by
\begin{equation}
(\zeta_{Q})_{V} := Q(\flat(V) \hookrightarrow V).
\end{equation}
Every presheaf $Q \in \Vboldhat$ is a $j$-sheaf if and only if 
$\zeta_{Q}$ is an isomorphism.
In particular, $\zeta_{\flat^{*}Q}$ is always isomorphic;
that is, $\flat^{*} Q$ is a $j$-sheaf for any presheaf $Q$.
In fact, $\flat^{*}:\Vboldhat \to \Sh_{j} \Vboldhat$,
where $\Sh_{j} \Vboldhat$ denotes the topos of $j$-sheaves, 
is the associated sheaf functor,
and $\zeta$ is the unit of the adjunction $\flat^{* } \dashv i$,
where $i:\Sh_{j} \Vboldhat \hookrightarrow \Vboldhat$ is the inclusion.

In the topos $\Sh_{j}\Vboldhat$, truth-values of physical propositions are taken on the Heyting algebra 
$\Gamma \Omega_{j} \equiv \Hom(1, \Omega_{j})$ of 
global elements of the sub-object classifier $\Omega_{j}$ of $\Sh_{j} \Vboldhat$.
(Here, $1$ is the terminal object of $\Vboldhat$, 
which gives the one-point set $1(V) = \{*\}$ for each $V \in \Vbold$ .)
As is well-known, $\Omega_{j}$ is a specific sub-object of $\Omega$,
i.e., the equalizer of $\Omega \xrightarrow{1_{\Omega}} \Omega$ and
$\Omega \xrightarrow{j} \Omega$.
In the present case, it is given by
\begin{equation}
\Omega_{j}(V) := \{ \omega \in \Omega(V) \;|\; \forall V' \subseteqVbold V \;
( \flat(V') \in \omega \Rightarrow V' \in \omega  )\}.
\label{eq:sub-object classifier of quantization topos}
\end{equation}
For each $V \in \Vbold$, $\Omega_{j}(V)$ contains 
the sieve  $\tmath_{V}$ consisting of all sub-algebras of $V$ as the top element.
The truth morphism $\true_{j} \in \Gamma \Omega_{j}$ is, therefore, given by
\begin{equation}
(\true_{j})_{V}:= \tmath_{V}.
\end{equation}

Let $\Sigma \in \Vboldhat$ be the spectral presheaf; that is,
it assigns to each context $V \in \Vbold$ the Gel'fand space $\Sigma(V)$ on $V$.
As physical propositions are represented by clopen sub-presheaves of $\Sigma$ in the theory of D\"{o}ring and Isham, 
we assume that they are represented by clopen sub-sheaves of the spectral sheaf $\flat^{*} \Sigma$. 
The space $\Sub_{j\, \cl} (\flat^{*}\Sigma)$ of $j$-sheaf propositions is represented in $\Sh_{j} \Vbold$ by
the $j$-sheaf $\Pmathbb_{j \, \cl} (\flat^{*} \Sigma)$ that is defined by 
\begin{equation}
(\Pmathbb_{j \, \cl} (\flat^{*} \Sigma))(V)  :=
\Sub_{j \, \cl}(\flat^{*} (\Sigma_{\downarrow V})) 
\label{eq:power object 1b}
\end{equation}
and
\begin{equation}
(\Pmathbb_{j \, \cl} (\flat^{*} \Sigma))(V' \hookrightarrow V) (S) := \flat^{*}(S_{\downarrow V'}).
\label{eq:power object' 2}
\end{equation}

Every physical proposition $P \in\Sub_{j \, \cl}(\flat^{*} \Sigma)$ has its name $\lceil P \rceil \in \Gamma \Pmathbb_{j \, \cl} (\flat^{*} \Sigma)$ 
defined by 
\begin{equation}
\lceil P \rceil _{V} := \flat^{*}(P_{\downarrow V}) \in \Sub_{j \, \cl}(\flat^{*}(\Sigma_{\downarrow V})).
\end{equation}

If we are given a truth sheaf $\Tmathbb_{j}$, which is a sub-sheaf of $\Pmathbb_{j \, \cl} (\flat^{*} \Sigma)$
that specifies truth propositions,
every physical proposition $P \in \Sub_{j \, \cl}(\flat^{*} \Sigma)$ is assigned a truth-value
$\nu_{j}(P;\Tmathbb_{j}) \in \Gamma \Omega_{j}$ via the diagram
\begin{equation}
\xymatrix{
&& \Tmathbb_{j} \ar [rr] ^{!} \ar @{>->} [dd] && 1 \ar @{>->} [dd] ^{\true_{j}} \\
&&&& \\
1 \ar @{>->} [rr] ^(0.4){\lceil P \rceil} && \Pmathbb_{j \, \cl}(\flat^{*} \Sigma)  \ar [rr] _{\tau_{j}} && \Omega_{j} \\
}
\label{eq:pullback j}
\end{equation}
where, $\tau_{j}$ is the characteristic morphism of $\Tmathbb_{j} \rightarrowtail  \Pmathbb_{j \, \cl}(\flat^{*} \Sigma) $,
which makes the square a pullback.
Thus, we have
\begin{eqnarray}
\nu_{j}(P;\Tmathbb_{j}) _{V}
& = & 
\tau_{i} \circ \lceil P \rceil \nonumber\\
& = &
\{
V' \subseteqVbold V 
\;|\;
\lceil P \rceil _{V'} \in \Tmathbb_{j}(V') 
 \}.
 \label{eq:truth-value j}
\end{eqnarray}

In ordinary quantum theory, every physical proposition $A \in \Delta$,
which means that the value of a physical quantity $A$ is in a Borel set $\Delta$,
is represented by a projection operator, which we denote by $\Ehat[A \in \Delta]$.
To represent such a proposition in $\Sh_{j} \Vboldhat$,
we introduce the (outer) daseinization operator $\delta_{j}$.
This is 
a $j$-sheaf counterpart of the daseinization operator $\delta$ introduced by D\"{o}ring and Isham \cite{DI08a,DI08b},
and is given as a global element of the outer sheaf \cite{N14} defined by
\begin{eqnarray}
(\delta_{j} (\Ehat))_{V} & :=  & (\delta (\Ehat))_{\flat(V)} \nonumber\\
& = & 
\bigwedge \{ \alphahat \in \Pcal(\flat(V)) 
\;|\; 
\Ehat \preceq \alphahat \}.
\label{eq:daseinization projection}
\end{eqnarray}
As is well-known, for each $V \in \Vbold$,
there exists a bijection between  the set $\Pcal(V)$ of projection operators in $V$
and the family $\Sub_{\cl}(\Sigma(V))$ of clopen subsets of $\Sigma(V)$ as
\begin{equation}
\Pcal(V) \xrightarrow{\sim} \Sub_{\cl}(\Sigma(V));\Phat \mapsto \{\sigma \in \Sigma(V) \;|\; \sigma(\Phat) = 1\}.
\label{eq:projection-clopen bijection}
\end{equation}
By the aid of this correspondence,
we can identify $\delta_{j}(\Ehat)$ with a clopen sub-sheaf of the spectral sheaf $\flat^{*} \Sigma$
\begin{eqnarray}
(\delta_{j}(\Ehat) )(V) 
&=&
\{ \sigma \in (\flat^{*} \Sigma)(V) \,|\, 
\sigma ((\delta_{j}(\Ehat))_{V})= 1\} \nonumber\\
& = &
\{ \sigma \in \Sigma(\flat(V)) \,|\, 
\sigma ((\delta(\Ehat))_{\flat(V)})= 1\} .
\label{eq:daseinization spectral sheaf}
\end{eqnarray} 

Let $\rho$ be a density matrix and $r \in [0,1]$.
D\"{o}ring and Isham \cite{DI12} defines generalized truth objects $\Tmathbbrhor$,
the global elements of which represent propositions that are only true with probability at least $r$ in the state $\rho$. 
Similarly, we define their $j$-sheaf counterparts $\Tmathbbrhor_{j} \in \Sub_{j\, \cl}(\flat^{*}\Sigma)$ by
\begin{equation}
\Tmathbbrhor_{j}(V)
:= \{
S \in \Sub_{j \, \cl}(\flat^{*}(\Sigma_{\downarrow V})) \;|\;
\tr(\rho (\Phat_{S(V)})) \ge r
\},
\label{eq:truth sheaf rho r}
\end{equation}
where $\Phat_{S(V)} \in \Pcal(V)$ is the projection operator given by $S(V)$ via the inverse of (\ref{eq:projection-clopen bijection}).

The truth-value $\nu_{j}(A \in \Delta;\Tmathbbrhor_{j}) $ of a proposition $A \in \Delta$ represented by a projection $\Ehat[A \in \Delta]$ under the physical state $\rho$ is 
given by
\begin{eqnarray}
\nu_{j}(A \in \Delta;\Tmathbbrhor_{j}) _{V}
& = &
\{ V' \subseteqVbold V
\;|\;
\lceil \delta_{j}(\Ehat[A \in \Delta]) \rceil_{V'} \in \Tmathbbrhor_{j}(V')\} \nonumber\\
& = &
\{ V' \subseteqVbold V
\;|\; 
\tr(\rho (\delta_{j}(\Ehat[A \in \Delta]))_{V'}) \ge r \}.
\label{eq:relation}
\end{eqnarray}

\subsection{Discrimination of physical propositions}
\label{sec:separability}

The mathematical ingredients constituting $j$-sheaf topos quantum theory 
can be obtained by reducing those of presheaf-based quantum theory in 
the manner described by Nakayama \cite{N14}.
That is, the spaces of propositions, truth-values, and truth objects represented by $j$-sheaves
are obtained by coarse graining the presheaf-based ones in the manner described by Nakayama \cite{N14}.
(We give a brief summary in the appendix.)
Accordingly, sheaf-based theories have less representation power than the presheaf theory.
In order for the sheaf-based theories to work well,
they need to be able to discriminate different physical propositions on properties that are objects of concern;
such propositions should be represented by different $j$-sheaves
and should be assigned different truth-values. 
Can the present theory correctly discriminate propositions that should be discriminated?
Our answer to this question is that $j$-sheaf based quantum theory does work all right at least on $\flat$-selected observables.
\begin{prp} \label{prp:P1 P2}
Let $P_{1}$ and $P_{2}$ be $j$-sheaf propositions;
that is,  $P_{1}$, $P_{2} \in \Sub_{j \, \cl} (\flat^{*}\Sigma)$.
If  $
\nu_{j}(P_{1}; \Tmathbb_{j}^{\rho}) = \nu_{j}(P_{2}; \Tmathbb_{j}^{\rho}) 
$
for all density matrices $\rho$,
we have
$P_{1} = P_{2}$.
Here, $\Tmathbb^{\rho}_{j} \equiv \Tmathbb^{\rho \, , 1}_{j}$.
\end{prp}
Proof.
Suppose that $P_{1} \ne P_{2}$.
Then, there exists $V \in \Vbold$ ($V \ne \Cmathbb \Ihat$)
such that 
\begin{equation}
P_{1}(V) = P_{1}(\flat(V)) \ne P_{2}(\flat(V)) = P_{2}(V),
\label{eq:Phat1 Phat2}
\end{equation}
hence,
\begin{equation}
\Phat_{P_{1}(V)} = \Phat_{P_{1}(\flat(V)) }\ne \Phat_{P_{2}(\flat(V))} = \Phat_{P_{2}(V)}.
\label{eq:P1 ne P2}
\end{equation}

Since $\Phat_{S_{1}(V)}$, $\Phat_{P_{2}(V)} \in \Pcal(\flat(V))$,
they are commutative.
Therefore, (\ref{eq:P1 ne P2}) implies 
\begin{equation}
\Phat_{P_{1}(V)} \perp \Phat_{P_{2}(V)} \quad \mbox{or} \quad 
\mbox{(say)} \; \Phat_{P_{1}(V)} \prec \Phat_{P_{2}(V)} ,
\end{equation}
where $\prec$ is the proper inequality sign excluding $=$.
In the both cases, we can take a state vector $\varphivec$
such that 
\begin{equation}
\Phat_{P_{2}(V)}\varphivec = \varphivec
\quad \mbox{and}\quad
\Phat_{P_{1}(V)}\varphivec = 0.
\end{equation}
Then, we have
\begin{equation}
\nu_{j}(P_{2};\Tmathbb^{\varphipro}_{j})_{V} = \tmath_{V},
\end{equation}
whereas
\begin{equation}
V \not\in \nu_{j}(P_{1};\Tmathbb^{\varphipro}_{j})_{V},
\end{equation}
hence, $\nu_{j}(P_{1};\Tmathbb^{\varphipro}_{j}) \ne \nu_{j}(P_{2};\Tmathbb^{\varphipro}_{j})$.
\qed
Consequently, whenever propositions $P_{1}$ and $P_{2}$ are represented by different $j$-sheaves,
we can always give a density matrix $\rho$ that assigns to them different truth values.
From proposition \ref{prp:P1 P2},
we can verify that the $j$-sheaf theory well separates physical propositions represented by $\flat$-selected projection operators.
\begin{thm}
Let $\Ehat_{1}$ and $\Ehat_{2}$ be $\flat$-selected projection operators.
Then, the following conditions are equivalent.
\begin{description}
\item[(i)] $\Ehat_{1} = \Ehat_{2}$.
\item[(ii)] $\delta_{j}(\Ehat_{1} )= \delta_{j}(\Ehat_{2})$.
\item[(iii)] For all density matrices $\rho$,
$\nu_{j}(\delta_{j}(\Ehat_{1}); \Tmathbb^{\rho}_{j}) = \nu_{j}(\delta_{j}(\Ehat_{2}); \Tmathbb^{\rho}_{j})$.
\end{description}
\end{thm}
Proof.
It is obvious that  (i) $\Rightarrow$ (ii) $\Rightarrow$ (iii) holds.
Also, (iii) $\Rightarrow $ (ii) is clear from Proposition \ref{prp:P1 P2}.
To show (ii) $\Rightarrow$ (i),
we note that (ii) implies that $\Ehat_{1}$ and $\Ehat_{2}$ are commutative.
In fact, if this is not the case,
there exist contexts $V_{1}$ and $V_{2} \in \Vbold$ such that $\Ehat_{1} \in \flat(V_{1})$ 
($\Ehat_{2} \not \in V_{1}$) and $\Ehat_{2} \in \flat(V_{2})$ ($\Ehat_{1} \not \in V_{2}$) ,
which lead us to a contradiction
\begin{equation}
\Ehat_{1} \prec \delta_{j}(\Ehat_{1})_{V_{2}} = \delta_{j}(\Ehat_{2})_{V_{2}} = \Ehat_{2}
\end{equation}
and
\begin{equation}
\Ehat_{2} \prec \delta_{j}(\Ehat_{2})_{V_{1}} = \delta_{j}(\Ehat_{1})_{V_{1}} = \Ehat_{1}.
\end{equation}
Thus, $\Ehat_{1}$ and $\Ehat_{2}$ have to commute,
hence, e.g., for $V \equiv \{\Ehat_{1},\,\Ehat_{2}\}''$,
\begin{equation}
\Ehat_{1} = \delta_{j}(\Ehat_{1})_{V} = \delta_{j}(\Ehat_{2})_{V} = \Ehat_{2}.
\end{equation}
\qed

In particular, 
if $A$ is a $\flat$-selected observable,
the projection operator $\Ehat[A \in \Delta]$ is also $\flat$-selected.
Thus, physical propositions $A \in \Delta$ on $\flat$-selected observables are well discriminated
in the $j$-sheaf based topos quantum theory.


\section{Measures on the spectral sheaf and quantum theory on $j$-sheaves}
\label{sec:probability measure}
\subsection{Measures on the spectral sheaf}

In the presheaf-based topos approach, quantum probabilities can be reproduced via the notion of measures on the spectral presheaf $\Sigma$ \cite{D09,DI12}.
D\"{o}ring \cite{D09} defined the measures as follows.
First, define a presheaf  $[0,1]^{\succeq} \in \Vboldhat$ of order-reserving maps by
\begin{eqnarray}
[0,1]^{\succeq} (V) 
& := &
\Hom^{\succeq}(\downarrow V, [0,1]) \nonumber\\
& \equiv &
\{
\downarrow V \xrightarrow{f} [0,1]
 \;|\; \forall \;V',\,V'' \in  \downarrow V \nonumber\\
 & &\qquad\qquad \qquad  (V'' \subseteqVbold V' \Rightarrow f(V'') \ge f(V'))\},
\end{eqnarray}
and
\begin{equation}
[0,1]^{\succeq} (V' \hookrightarrow V)(f) = f|_{\downarrow V'}. 
\end{equation}
Then, a measure $\mu$ on $\Sigma$ is defined as a map 
\begin{equation}
\mu: \Sub_{\cl}(\Sigma) \to  \Gamma [0,1]^{\succeq}\;;
S \mapsto  (\mu(S)_{V})_{V \in \Vbold} ,
\end{equation}
such that
\begin{equation}
(\mu(\Sigma))_{V}(V') = 1
\label{eq:mu(Sigma)}
\end{equation}
and for all $S_{1},\;S_{2} \in \Sub_{\cl}(\Sigma)$,
\begin{equation}
\mu(S_{1} \vee S_{2}) + \mu(S_{1} \wedge S_{2}) = \mu(S_{1}) + \mu(S_{2}).
\label{eq:mu additivity}
\end{equation}
Futhermore, for each $V \in \Vbold$,
$(\mu(S))_{V}$ is assumed to be determined by $S(V) \in \Sub_{\cl} (\Sigma(V))$;
that is,
there exists a unique $ (\mu(S(V)))_{V \in \Vbold} \in \Hom^{\succeq}(\Vbold, [0,1])$
such that, for $V' \in \downarrow V$, 
\begin{equation}
(\mu(S))_{V}(V') =\mu(S(V')).
\label{eq:mu locality}
\end{equation}

Every quantum state represented by a density matrix $\rho$ defines 
the associated measure $\mu^{\rho} \in \Hom^{\succeq}(\Vbold,[0,1])$ by, 
for each $S \in  \Sub_{\cl} \Sigma$, 
\begin{equation}
\mu^{\rho}(S(V)) = \tr (\rho \Phat_{S(V)}).
\label{eq:mu^rho}
\end{equation}
Conversely, if the von Neumann algebra $\Bcal(\Hcal)$ of bounded operators on $\Hcal$ has no direct summand of type $I_{2}$,
every measure $\mu$ on $\Sigma$ uniquely determines a density matrix $\rho_{\mu}$
that gives $\mu$ via (\ref{eq:mu^rho}).

D\"{o}ring \cite{D09} showed, furthermore,  
that the quantum probability $\Prob(A \in \Delta)$ with which the physical proposition $A \in \Delta$ is true can be reproduced from $\mu$ as 
\begin{equation}
\Prob(A \in \Delta) = \min_{V \in \Vbold} \mu^{\rho}(\delta(\Ehat[A \in \Delta])(V)).
\end{equation}

In a similar way,
we can define a measure $\mu_{j}$ on the spectral sheaf $\flat^{*} \Sigma$.
To do so, we define a sub-presheaf $\Hom^{\succeq}_{\flat}$ of $[0,1]^{\succeq}$ by
\begin{equation}
\Hom^{\succeq}_{\flat}(V) := \{ h \in [0,1]^{\succeq}(V) \;|\; \forall V' \in \downarrow V \;(h(\flat(V')) = h(V')) \},
\end{equation}
and furthermore, define a $j$-sheaf $[0,1]^{\succeq}_{j} $ by
\begin{equation}
[0,1]^{\succeq}_{j} := \flat^{*} (\Hom^{\succeq}_{\flat}).
\end{equation}
Then, it holds that
\begin{equation}
\Gamma [0,1]^{\succeq}_{j} \simeq 
\Hom^{\succeq}(\flat(\Vbold),[0,1]) ,
\end{equation}
where the bijective correspondence is given as follows.
For every $\gamma \in \Gamma [0,1]^{\succeq}_{j} $,
$h_{\gamma} \in \Hom^{\succeq}(\flat(\Vbold),[0,1])$ is defined by
\begin{equation}
h_{\gamma}(\flat(V)) = \gamma_{\flat(V)}(\flat(V)) = \gamma_{V}(V) ,
\label{eq:gamma to h}
\end{equation}
and conversely,
for $h \in \Hom^{\succeq}(\flat(\Vbold), [0,1])$, $\gamma_{h} \in \Gamma[0,1]^{\succeq}_{j}$ is given by
\begin{equation}
(\gamma_{h})_{V}(V') := h(\flat(V')).
\end{equation}

Following (\ref{eq:mu(Sigma)})-(\ref{eq:mu locality}),
we define a measure $\mu_{j}$ on the spectral sheaf $\flat^{*}\Sigma$ as 
a map 
\begin{equation}
\mu_{j}: \Sub_{j \, \cl}(\flat^{*} \Sigma) \to \Gamma[0,1]^{\succeq}_{j} \simeq \Hom^{\succeq}(\flat(\Vbold),[0,1]);S \mapsto (\mu_{j}(S(\flat(V))))_{V \in \Vbold}
\end{equation}
such that, for all $V' \in \downarrow V$
\begin{equation}
(\mu_{j}(\flat^{*}\Sigma))_{V}(V') = 1
\end{equation}
and for all $S_{1},\;S_{2} \in \Sub_{j \, \cl}(\flat^{*}\Sigma)$,
\begin{equation}
\mu_{j}(S_{1} \vee S_{2}) + \mu_{j}(S_{1} \wedge S_{2}) = \mu_{j}(S_{1}) + \mu_{j}(S_{2}).
\end{equation}
Note that $\mu_{j}(S) \in \Gamma [0,1]^{\succeq}_{j}$ and $\mu_{j}(S(-)) \in \Hom^{\succeq}(\flat(\Vbold),[0,1])$ are related each other as
\begin{equation}
(\mu_{j}(S))_{V}(V') = \mu_{j}(S(\flat(V'))).
\end{equation}

For every density matrix $\rho$,
the associated measure $\mu^{\rho}_{j}$ is defined by 
\begin{equation}
\mu^{\rho}_{j}(S(\flat(V))) :=  \tr (\rho \Phat_{S(\flat(V))}) = \tr (\rho \Phat_{S(V)}) .
\end{equation}

Differently from $\mu^{\rho}$,
$\mu^{\rho}_{j}$ does not give correct quantum probabilities for all physical propositions of type of $A \in \Delta$.
In fact, since we have 
\begin{eqnarray}
\{ \delta_{j}(\Ehat(A \in \Delta))_{V} \}_{V \in \Vbold}
& = &
\{ \delta(\Ehat(A \in \Delta))_{\flat(V)} \}_{V \in \Vbold} \nonumber\\
& \subseteq &
\{ \delta(\Ehat(A \in \Delta))_{V} \}_{V \in \Vbold},
\end{eqnarray}
the only thing we can say is that
\begin{equation}
\Prob (A \in \Delta) \le \min_{V \in \Vbold} \tr (\rho\delta_{j}(\Ehat(A \in \Delta))_{V}) = \min_{V \in \Vbold} \mu^{\rho}_{j} (S(\flat(V))).
\label{eq:probability sheaf}
\end{equation}
However,
if $A$ is $\flat$-selected, namely $A \in \flat(V)$ for some $V \in \Vbold$,
then $\Ehat(A \in \Delta) \in \Pcal(V)$, hence, $\delta_{j}(\Ehat(A \in \Delta)) )_{V} =\Ehat(A \in \Delta) $.
Thus, we obtain the following result:
\begin{prp}
If $A$ is  a $\flat$-selected physical quantity,
the probability with which the proposition $A \in \Delta$ is true is correctly given by
\begin{equation}
\Prob (A \in \Delta) =  \min_{V \in \Vbold} (\mu^{\rho}_{j} (\delta_{j}(\Ehat[A \in \Delta])))_{V}.
\end{equation}
\end{prp}
As mentioned in section \ref{sec:separability},
the topos quantum theory on $j$-sheaves works well for $\flat$-selected physical quantities.
This is the case also for calculation of quantum probabilities.

\subsection{Measure morphisms and truth-value valuation}

\def\mutilde{\tilde{\mu}}

If we are given a measure  $\mu_{j}$ on $\flat^{*}\Sigma$,
we can always relate it to truth-value valuation in the sheaf topos $\Sh_{j}\Vboldhat$.
To see this, we show that $\mu_{j}$ can be   internalized to $\Sh_{j} \Vbold$.
For every measure $\mu_{j}$,
we define a morphism $\Pmathbb_{j \, \cl}(\flat^{*} \Sigma) \xrightarrow{\mutilde_{j}} [0,1]^{\succeq}_{j}$ by
\begin{equation}
(\mutilde_{j})_{V}\;:\; \Sub_{j \, \cl}(\flat^{*}( \Sigma_{\downarrow V})) \to [0,1]^{\succeq}_{j}(V) \;;\; S \mapsto (\mu_{j}(S))_{V} .
\end{equation}
In fact, since $\mutilde_{j}$ satisfies the naturality condition as
\begin{eqnarray}
(\mutilde_{j})_{V'}(\flat^{*}(S_{\downarrow V'}))  (W)
& = &
(\mu_{j}(\flat^{*}(S_{\downarrow V'})))_{W} \nonumber\\
& = &
\mu_{j}(S_{\downarrow V'}(\flat(W))) \nonumber\\
& = &
\mu_{j}(S(\flat(W))) \nonumber\\
& = &
((\mutilde_{j})_{V}(S))(W),
\label{eq:mu morphism}
\end{eqnarray}
it is really a morphism in $\Sh_{j} \Vboldhat$.
Next, for each $r \in [0,1]$, 
let us define a morphism
$[0,1]^{\succeq}_{j} \xrightarrow{\lambda_{j}^{r}} \Omega_{j}$ by
\begin{equation}
(\lambda_{j}^{r})_{V}:[0,1]^{\succeq}_{j}(V) \to \Omega_{j}(V);
h \mapsto 
\omega_{h} := \{V' \subseteqVbold V \;|\; h(V') \ge r \},
\label{eq:lambda morphism}
\end{equation}
which is truly a morphism as can be easily shown.

Any density matrix $\rho$ induces a measure $\mu_{j}^{\rho}$
and a truth sheaves $\Tmathbb_{j}^{\rho,\,r}$ with $r \in [0,1]$.
They are related each other via (\ref{eq:truth sheaf rho r}) and (\ref{eq:mu^rho}).
This relation can be expressed in $\Sh_{j} \Vboldhat$ via $\tilde{\mu_{j}^{\rho}}$.
We can easily prove the following statement,
which is just a restatement of definition (\ref{eq:truth sheaf rho r}). 
\begin{thm}
Let $\rho$ be a density matrix and $r \in [0,1]$.
Then, the characteristic morphism $\Pmathbb_{j \, \cl}(\flat^{*}\Sigma) \xrightarrow{\tau_{j}^{\rho,\, r}} \Omega_{j}$ of the truth sheaf $\Tmathbb_{j}^{\rho,\,r}$ 
is factored through $[0,1]^{\succeq}_{j}$ as
\begin{equation}
\xymatrix{
\Pmathbb_{j \, \cl}(\flat^{*}\Sigma) \ar [rr] ^{\tau_{j}^{\rho,\,r}} \ar [rdd] _{\mutilde_{j}^{\rho}}&& \Omega_{j} \\
&& \\
&   [0,1]^{\succeq}_{j} \ar [ruu] _{\lambda_{j}^{r}} & \\
}
\end{equation}
\end{thm}

Let us define a global element  $h_{\max}$ of $ [0,1]^{\succeq}_{j}$ by
\begin{equation}
(h_{\max})_{V}(V') = 1.
\end{equation}
Then, every measure $\mu_{j}$ induces a canonical truth sheaf $\Tmathbb_{j}^{\mu_{j}}$
as a pullback of $h_{\max}$ along $\mutilde_{j}$;
Namely, $\Tmathbb_{j}^{\mu_{j}}$ makes the diagram
\begin{equation}
\xymatrix{
\Tmathbb_{j}^{\mu_{j}} \ar [rr] ^{!} \ar @{>->} [dd] && 1 \ar @{>->} [dd] ^{h_{\max}} \\
&&  \\
\Pmathbb_{j\, \cl}(\flat^{*}\Sigma) \ar [rr] _{\mutilde_{j}} &&  [0,1]^{\succeq}_{j} \\
}
\end{equation}
pullback, hence, is given as
\begin{eqnarray}
\Tmathbb^{\mu_{j}}(V)
& = &
\{
S \in \Sub_{j \, \cl}(\flat^{*}(\Sigma_{\downarrow V}))
\;|\;
((\mutilde_{j})_{V}(S) )(V') = 1
\} \nonumber\\
& = &
\{
S \in \Sub_{j \, \cl}(\flat^{*}(\Sigma_{\downarrow V}))
\;|\;
\mu_{j}(S(\flat(V'))) = 1
\}.
\end{eqnarray}
In particular, we have

\begin{equation}
\Tmathbb_{j}^{\rho,\,1} = \Tmathbb_{j}^{\mu_{j}^{\rho}}.
\end{equation}

\begin{prp}
The $j$-sheaf $\Tmathbb_{j}^{\mu_{j}}$ defined above is a truth sheaf.
\end{prp}
Proof.
What we should show is that $\Tmathbb^{\mu_{j}}(V)$ is a filter for any $V \in \Vbold$.
Let $S \in \Tmathbb^{\mu_{j}}(V)$ and $S \subseteq S' \in \Sub_{j \, \cl}(\flat^{*} (\Sigma_{\downarrow V}))$.
Then, since we have $\mu_{j}(S) \le \mu_{j}(S')$,
it follows that $\mu_{j}(S'(\flat(V'))) = 1$ for all $V' \subseteqVbold V$, which implies that $S' \in \Tmathbb^{\mu_{j}}(V)$ .

Let $S \in \Tmathbb^{\mu_{j}}(V)$  and $S' \in \Tmathbb^{\mu_{j}}(V)$.
Then, we have
\begin{equation}
\mu_{j}((S \wedge S')(V')) = \mu_{j}(S(V')) +\mu_{j}(S'(V')) -\mu_{j}((S \vee S')(V')) =1,
\end{equation}
which means that $S \wedge S' \in \Tmathbb^{\mu_{j}}(V)$.
\qed
We can easily verify the following statement.
\begin{thm}
Let $\tau_{j}^{\mu_{j}}$ be the characteristic morphism of $\Tmathbb_{j}^{\mu_{j}}$ defined above.
Then it is factored through $  [0,1]^{\succeq}_{j}$ as
\begin{equation}
\xymatrix{
\Pmathbb_{j \, \cl}(\flat^{*}\Sigma) \ar [rr] ^{\tau_{j}^{\mu_{j}}} \ar [rdd] _{\mutilde_{j}}&& \Omega_{j} \\
&& \\
&   [0,1]^{\succeq}_{j} \ar [ruu] _{\lambda_{j}} & \\
}
\end{equation}
where $\lambda_{j} := \lambda_{j}^{1}$.
\end{thm}
Consequently, if we are given a measure $\mu^{j}$,
we can assign to each proposition $P \in \Sub_{j \, \cl}(\flat^{*} \Sigma)$
a truth-value $\nu_{j}^{\mu_{j}}(P)$ as
\begin{equation}
\nu_{j}^{\mu_{j}}(P) 
:=
\nu_{j}(P;\Tmathbb_{j}^{\mu_{j}}) 
=
\lambda_{j} \circ \mutilde_{j} \circ \lceil P \rceil_{j}.
\end{equation}

Finally, we note relations between measures on the spectral presheaf $\Sigma$ and a measure $\mu_{j}$ on the spectral sheaf $\flat^{*}\Sigma$.
Obviously, every measure $\mu$ on $\Sigma$ defines a measure $\mu_{j}$ on $\flat^{*}\Sigma$ by
\begin{equation}
\mu_{j}(S(\flat(V))) := \mu(S(\flat(V))).
\end{equation}
Conversely,
for every $\mu_{j}$, 
any measure $\mu$ on $\Sigma$ that satisfies
\begin{equation}
\mu(S(\flat(V))) = \mu_{j}(\flat^{*}S(\flat(V)))
\end{equation}
for $S \in \Sub_{\cl}(\Sigma)$ defines a truth presheaf $\Tmathbb^{\mu}$
that satisfies (\ref{eq:translation of truth object}),
namely,  a translation of $\Tmathbb_{j}^{\mu_{j}}$.
One of such examples is given by
\begin{equation}
\mu(S(V)) := \mu_{j}(\flat^{*}S (\flat(V))),
\label{eq:max mu}
\end{equation}
This is the largest among the possible measures that gives $\mu_{j}$,
and induces the largest tranlation (\ref{eq:jmathboldmax}) of $\Tmathbb_{j}$.
Consequently, if the von Neumann algebra $\Bcal(\Hcal)$ has no direct summand of type $I_{2}$,
any $\mu_{j}$ (hence, $\Tmathbb_{j}^{\mu_{j}}$) can be reproduced as a restriction of 
a measure $\mu^{\rho}$ on $\Sigma$ given by a suitably defined density matrix $\rho$.

\section{Grothendieck topos with quantum probabilities as truth-values}
\label{sec:topology for quantum probability}

D\"{o}ring and Isham \cite{DI12} defines a sheaf topos 
in which probabilities can be regarded as truth-values, i.e.,
global elements of the sub-object classifier.
They defines a topological space $(0,1)_{L}$ whose open sets are open intervals $(0,r)$ ($r \in [0,1]$). 
Then, the  space $I \equiv [0,1]$ of probabilities can be regarded as the family $\Ocal((0,1)_{L})$ of the open sets
via the bijection $I \to \Ocal((0,1)_{L}) \,; \, r \mapsto (0,r)$.
Let $\Sh((0,1)_{L})$ be the sheaf topos on $(0,1)_{L}$.
Its sub-object classifier $\Omega^{(0,1)}$ is given as
\begin{equation}
\Omega^{(0,1)}((0,r)) = \{(0,r') \;|\; 0 \le r' \le r\} \simeq [0,r].
\end{equation}
Then, each probability $p \in I$ is identified with a global element of $\Omega^{(0,1)}$
by an injection $\ell: \Ocal((0,1)_{L}) \to \Gamma \Omega^{(0,1)}$ defined by
\begin{equation}
(\ell(p))_{(0,r)} := 
\{(0,r') \;|\; r' = \min \{p,r\} \}.
\end{equation}

What we do in this section is the same,
though we would rather proceed in a somewhat different way for later convenience.
First, note that $I$ can be regarded as a category
with probabilities $r \in I$ as objects
and morphisms $r' \hookrightarrow r$ that are defined if and only if $r' \le r$.  
We start with the topos $\Ihat \equiv \Set^{I^{\op}}$ of presheaves 
on the space $I$ of probabilities.
We introduce a Grothendieck topology on $I$ 
that induces a topos whose sub-object classifier is isomorphic to $\Omega^{(0,1)}$.

The sub-object classifier $\Omega_{\Ihat}$ of $\Ihat$ is obviously given by
\begin{equation}
\Omega_{\Ihat}(r) = \{\downarrow r' \;|\; r' \in [0,r]\} \cup \{ \downarrow \mathring{r'} \;|\; r' \in [0,r]\},
\end{equation}
where 
\begin{equation}
\downarrow r  :=  [0,r]
\quad\mbox{and}\quad
\downarrow \mathring{r}  :=  [0,r),
\end{equation}
and, in particular, 
\begin{equation}
\downarrow 0 = [0,0] = \{0\}
\quad\mbox{and}\quad
\downarrow \mathring{0} = (0,0) = \emptyset.
\end{equation}

The sub-object classifier $\Omega_{\prob}$ of the sheaf topos  we are seeking
should be a sub-object of $\Omega_{\Ihat}$ that satisfies
$\Omega_{\prob}(r) \simeq \Omega^{(0,1)}((0,r))$.
So, we assume that  $\Omega_{\prob}$ is given by
\begin{equation}
\Omega_{\prob}(r)  = \{\downarrow r' \;|\; r' \in [0,r]\},
\label{eq:OmegaProb}
\end{equation}
since $\Omega_{\prob}(r)$ need to contain $\tmath_{r} = \downarrow r$.

If  $\Omega_{\prob}$ is a sub-object classifier of a sheaf topos,
the corresponding Lawvere-Tierney topology is a morphism $\Omega_{\Ihat} \xrightarrow{j_{\prob}} \Omega_{\Ihat}$ that makes the diagram
\begin{equation}
\xymatrix{
\Omega_{\prob} \ar @{>->} [r] & \Omega_{\Ihat} \ar @<1mm> [r] ^{1_{\Omega_{\Ihat}}}\ar @<-1mm> [r]_{j_{\prob}} & \Omega_{\Ihat} \\
}
\end{equation}
an equalizer diagram.
Such a $j_{\prob}$ is given as,
for every $\varpi \in \Omega_{\Ihat}(r)$,
\begin{equation}
(j_{\prob})_{r}(\varpi) := \downarrow \sup \varpi.
\end{equation}
It is easy to show that the morphism $j_{\prob}$ is really a Lawvere-Tierney topology on $\Ihat$.
Thus, $\Omega_{\prob}$ defned by (\ref{eq:OmegaProb}) is a sub-object classifier of
the sheaf topos $\Sh_{\prob} \Ihat$ induced by $j_{\prob}$.
The probabilities are identified with global elements of $\Omega_{\prob}$ by the injection $\ell':I \to \Gamma \Omega_{\prob};p \mapsto \ell'(p)$
defined by
\begin{equation}
(\ell'(p))_{r} := \{\downarrow r' \;|\; r' = \min\{p,r\}\}.
\end{equation}

For convenience in Section \ref{sec:ibold-sheaf topos}, 
we describe the sheaf topos $\Sh_{\prob} \Ihat$ more fully in the following.

The Grothendieck topology $J_{\prob}$ corresponding to $j_{\prob}$
is a sub-object of $\Omega_{\prob}$ whose characteristic morphism is  $j_{\prob}$.
Therefore, it is given as
\begin{equation}
J_{\prob}(r) := \{ \downarrow r , \; \downarrow \mathring{r} \}.
\end{equation} 
The closure operator corresponding to $j_{\prob}$ is given as,
for a presheaf $Q \in \Ihat$ and a sub-presheaf $S \in \Sub(Q)$,
\begin{equation}
\overline{S}(r) := 
\{
q \in Q(r) \;|\;
\forall r' \in \downarrow \mathring{r} \; (q|_{r'} \in S(r'))
\}.
\end{equation}

To describe $j_{\prob}$-sheaves,
we introduce a functor $a_{\prob}: \Ihat \to \Ihat$,
which is defined as, for each $Q \in \Ihat$ and for  $r \in I$, 
\begin{equation}
(a_{\prob}Q)(r)  := 
\varprojlim_{s \in \downarrow \mathring{r}} Q(s) 
= \Hom_{\Ihat}(1_{\downarrow \mathring{r}},Q)  ,
\end{equation}
for each $r' \hookrightarrow r \in \Mor(I)$,
\begin{equation}
(a_{\prob}Q)(r' \hookrightarrow r) := \Hom_{\Ihat}(1_{\downarrow \mathring{r'}} \rightarrowtail 1_{\downarrow \mathring{r}},Q),
\end{equation}
and for  $Q \xrightarrow{f} R \in \Mor(\Ihat)$,
\begin{equation}
(a_{\prob}Q)(f) := \Hom_{\Ihat}(1_{\downarrow \mathring{r}},f).
\end{equation}
In addition, we define a natural transformation $I \xrightarrow{\zeta^{\prob}} a_{\prob}$ by
\begin{equation}
(\zeta^{\prob}_{Q})_{r}:= \Hom_{\Ihat}(1_{\downarrow \mathring{r}} \rightarrowtail 1_{\downarrow r},Q):
Q(r)\simeq \Hom_{\Ihat}(1_{\downarrow r},Q)  \to \Hom_{\Ihat}(1_{\downarrow \mathring{r}},Q).
\end{equation}
As can be seen from the definition of $J_{\prob}$,
a presheaf $Q \in \Ihat$ is a $J_{\prob}$-sheaf (hence, a $j_{\prob}$-sheaf)
if and only if $\zeta^{\prob}_{Q}$ is a natural isomorphism.
Furthermore, for every presheaf $Q \in \Ihat$,
$a_{\prob} Q$ is a $j_{\prob}$-sheaf
because $\zeta^{\prob}_{a_{\prob}Q}$ is isomorphic
as is shown by the commutative diagram
\begin{equation}
\xymatrix{
\Hom_{\Ihat}(1_{\downarrow r}, a_{\prob}Q) \ar [rr] ^{(\zeta^{\prob}_{a_{\prob}Q})_{r}} \ar  [rrd] \ar [dd] _{\vertsim}&& \Hom_{\Ihat}(1_{\downarrow \mathring{r}},a_{\prob}Q) \ar  [d]^{\vertsim}\\
&& \varprojlim_{s \in \downarrow \mathring{r}} \Hom_{\Ihat}(1_{\downarrow s}, a_{\prob} Q) \ar [d] ^{\vertsim}\\
\Hom_{\Ihat}(1_{\downarrow \mathring{r}},Q) \ar _(0.45){\sim} [rr] && \varprojlim_{s \in \downarrow \mathring{r}} \Hom_{\Ihat}(1_{\downarrow \mathring{s}},Q) 
}
\end{equation}
In fact, the functor $a_{p}:\Ihat \to \Sh_{\prob} \Ihat$ is an associated sheaf functor,
and $\zeta^{\prob}$ is the unit of the adjunction $a_{\prob} \dashv i_{\prob}$,
where $i_{\prob}:\Sh_{\prob} \Ihat \hookrightarrow \Ihat$ is the inclusion.

\section{Sheaf-based topos quantum theory with quantum probabilities as truth-values}
\label{sec:ibold-sheaf topos}

Let $\Cbold := \Vbold \times I$.
We regard $\Cbold$ as a category consisting of pairs $(V,r) \in \Vbold \times I$ as  objects
and morphisms $(V',r') \hookrightarrow (V,r)$ given by the natural order relation $(V',r') \le_{\Cbold} (V,r)$,
which is defined 
if and only if $V' \subseteq_{\Vbold} V$ and $r' \le r$.
We write $\Cboldhat$ for the presheaf topos $\Set^{\Cbold^{\op}}$.
Also, we denote $\Omegabold$ for the sub-object-classifier of $\Cboldhat$.
That is, $\Omegabold(V,r)$ is the set of all sieves on $(V,r)$,
and  $\Omegabold((V',r')\hookrightarrow (V,r) )$ is defined as, for $\omegabold \in \Omegabold(V,r)$,
\begin{eqnarray}
\Omegabold((V',r')\hookrightarrow (V,r) )(\omegabold)
& = &
\{ (W,s) \in \omegabold
\;|\;
(W,s) \le_{\Cbold} (V',r')\} \nonumber\\
& = &
\omegabold \; \cap \downarrow (V',r') .
\end{eqnarray}

In the following, we construct a topos quantum theory on which both of the context-selection via $\flat$
and quantum probabilities as truth-values are reflected.
This is an extension of the theory of D\"{o}ring and Isham \cite{DI12}.
To do so, we construct a Grothendieck topology on $\Cbold$ from $J$ and $J_{\prob}$ first. 

We note that  $J$ can be generated by a coverage \cite{J02a} $K$ defined by
\begin{equation}
K(V) := \{ \downarrow \flat (V) \}.
\end{equation}
Namely, for each $\omega \in \Omega(V)$,
$\omega \in J(V)$ if and only if $\downarrow \flat(V) \subseteq \omega$.
Similarly,  $J_{\prob}$ can be generated by a coverage $K_{\prob}$ defined by
\begin{equation}
K_{\prob}(r) := \{  \downarrow \mathring{r} \}.
\end{equation}
We define $\Kbold$, a coverage on $\Cbold$, by
\begin{eqnarray}
\Kbold(V,r) &:=& 
 \{ \downarrow \flat(V) \times \downarrow \mathring{r}  \}\nonumber\\
& = &
\{
\{
(V',r') \in \Cbold \;|\; 
V' \in \downarrow \flat(V) \mbox{ and } r' \in \downarrow \mathring{r}
\}\}.
\end{eqnarray}
Hereafter we write $\downarrow (V, \mathring{r})$ for $ \downarrow V \times \downarrow \mathring{r}$.

By means of $\Kbold$, we define a set-valued map $\Jbold$ by
\begin{eqnarray}
\Jbold(V,r) 
& := &
\{\omegabold \in \Omegabold(V,r) \;|\; \downarrow (\flat(V), \mathring{r}) \subseteq \omegabold \} \nonumber\\
& = &
\{
\omegabold \in \Omegabold(V,r) \;|\; 
\forall r' \in \downarrow \mathring{r} \;((\flat(V),r') \in \omegabold)
\}.
\end{eqnarray}
It is easy to see that $\Jbold$ is a sub-object of $\Omegabold$,
and furthermore, a Grothendieck topology on $\Cbold$.

The Lawvere-Tierney topology $\jbold$ on $\Cboldhat$ corresponding to $\Jbold$ is defined by
\begin{eqnarray}
\jbold_{(V,r)}(\omegabold) & := &
\{
(V',r') \le_{\Cbold} (V,r)
\;|\;
\omegabold|_{(V',r')} \in \Jbold(V',r') 
\} \nonumber\\
& = &
\{
(V',r') \le_{\Cbold} (V,r)
\;|\;
\downarrow (\flat(V'),\mathring{r'}) \subseteq \omegabold 
\}.
\end{eqnarray}
Also, the corresponding closure operator is given as, for $\Sbold \in \Sub(\Qbold),$
\begin{equation}
\overline{\Sbold}(V,r)
=
\{
q \in \Qbold(V,r)
\;|\;
\forall s \in \downarrow \mathring{r} \;
(q|_{(\flat(V),s)} \in \Sbold(\flat(V),s))
\}.
\end{equation}

We write $\Sh_{\jbold} \Cboldhat$ for the sheaf topos induced by $\jbold$.
The sub-object classifier $\Omegabold_{\jbold}$ of $\Sh_{\jbold} \Cboldhat$ is given by
\begin{equation}
\Omegabold_{\jbold}(V,r) := \{ \omegabold \in \Omegabold(V,r) \;|\; \forall  (V',r') \le_{\Cbold} (V,r)
\; ( \downarrow (\flat(V'), \mathring{r'}) \subseteq \omegabold \Rightarrow (V',r') \in \omegabold ) \}.
\end{equation}
\def\flatbold{\boldsymbol{\flat}}
\def\bbold{\boldsymbol{b}}
\def\fbold{\boldsymbol{f}}
\def\qbold{\boldsymbol{q}}
\def\Rbold{\boldsymbol{R}}
\def\zetabold{\boldsymbol{\zeta}}
\def\pibold{\boldsymbol{\pi}}
\def\unitbold{\boldsymbol{1}}

\def\nubold{\boldsymbol{\nu}}
The sheafification functor associated with $\jbold$ is obtained from the composition of $a_{j}$ and $a_{\prob}$.
In the following, we describe this in a rigorous manner.

First, let us extend the context selector $\flat:\Vbold \to \Vbold$ to an endofunctor
$\flatbold:\Cbold \to \Cbold$, by
\begin{equation}
\flatbold(V,r):= (\flat(V),r)
\quad \mbox{and} \quad
\flatbold((V',r') \hookrightarrow (V,r)) :=(\flat(V'),r') \hookrightarrow (\flat(V),r)
\end{equation}
Then, the pullback functor $\flat^{*}$ is extended to $\flatbold^{*}:\Cboldhat \to \Cboldhat$ as
\begin{equation}
(\flatbold^{*} \Qbold)(V,r)= \Qbold(\flat(V),r) = ( \flat^{*}(\Qbold(-,r))) (V) .
\end{equation}
The morphism $1 \xrightarrow{\zeta} \flat^{*} \in \Mor(\Vboldhat)$ has its counterpart 
$\unitbold \xrightarrow{\zetabold^{\flat}} \flatbold^{*} \in \Mor(\Cboldhat)$ that is defined by,
for each $\Qbold \in \Cboldhat$ and $(V,r) \in \Cbold$,
\begin{equation}
(\zetabold^{\flat}_{\Qbold})_{(V,r)} := (\zeta_{\Qbold(-,r)})_{V}.
\end{equation}
Note that every presheaf $\Qbold(-,r)$ is a $j$-sheaf for an arbitrary $r \in I$ if and only if
$\zetabold^{\flat}_{\Qbold}$ is a natural isomorphism.
In particular, for every presheaf $\Qbold \in \Cboldhat$,
$(\flatbold^{*} \Qbold)(-,r)$ is a $j$-sheaf for any $r \in I$,
hence, $\zetabold^{\flat}_{\flatbold^{*}\Qbold}$ is always isomorphic. 

Next, we extend $a_{\prob}:\Ihat \to \Ihat$ to $\abold_{\prob}:\Cboldhat \to \Cboldhat$ by
\begin{eqnarray}
(\abold_{\prob} \Qbold)(V,r) 
&:=&  (a_{\prob}(\Qbold(V,-)))(r) \nonumber\\
& = &
\Hom_{\Ihat}(1_{\downarrow \mathring{r}} , \Qbold(V,-)) \nonumber\\
& \simeq &
\Hom_{\Cboldhat}(\unitbold_{\downarrow (V,\mathring{r})} , \Qbold) .
\end{eqnarray}
Here, the bijection from the second line to the third is given by
\begin{equation}
(q_{s})_{s \in \downarrow \mathring{r}} \mapsto (\Qbold(W \hookrightarrow V,s)(q_{s}))_{(W,s) \in \downarrow (V,\mathring{r})},
\label{eq:bijection from one to two}
\end{equation}
the inverse of  which is given by
\begin{equation}
 (q_{(W,s)})_{(W,s) \in \downarrow (V,\mathring{r})} \mapsto (q_{(V,s)})_{s \in \downarrow \mathring{r}} .
\end{equation}
As an extension of $I \xrightarrow{\zeta^{\prob}} a_{\prob} \in \Mor(\Ihat)$,
we define $\zetabold^{\prob}:\Ibold \to \abold_{\prob} \in \Mor(\Cboldhat)$ by
\begin{equation}
(\zetabold^{\prob}_{\Qbold})_{(V,r)} := (\zeta^{\prob}_{\Qbold(V,-)})_{r}.
\end{equation}
The presheaf $\Qbold(V,-)$ is $j_{\prob}$-sheaf
if and only if $\zetabold^{\prob}_{\Qbold}$ is a natural isomorphism.
Obviously, $\abold_{\prob}\Qbold(V,-)$ is always a $j_{\prob}$-sheaf, hence,
$\zetabold^{\prob}_{\abold \Qbold}$ is isomorphic.

Finaly, let us define $\abold:\Cboldhat \to \Cboldhat$ by
\begin{equation}
\abold :=   \flatbold^{*} \abold_{\prob},
\end{equation}
namely,
\begin{eqnarray}
(\abold \Qbold)(V,r)
& = &
((\flatbold^{*}\abold_{\prob}  )\Qbold)(V,r)   \nonumber\\
& = &
(\abold_{\prob}\Qbold)(\flat(V),r)\nonumber\\
&=&
\Hom(1_{\downarrow \mathring{r}},\Qbold(\flat(V),-)) \nonumber\\
& \simeq &
 \Hom_{\Cboldhat}(\unitbold_{\downarrow (\flat(V), \mathring{r})},\Qbold),
\end{eqnarray}
\begin{equation}
(\abold \Qbold)((V',r') \hookrightarrow  (V,r)) = 
\Hom_{\Cboldhat}(\unitbold_{\downarrow (\flat(V'), \mathring{r'})} \rightarrowtail \unitbold_{\downarrow (\flat(V), \mathring{r})},\Qbold),
\end{equation}
and for $\Qbold \xrightarrow{\fbold} \Rbold \in \Mor(\Cboldhat)$,
\begin{equation}
(\abold\fbold)_{(V,r)}
=
\Hom_{\Cboldhat}(\unitbold_{\downarrow (\flat(V), \mathring{r})},\fbold).
\end{equation}
Note that 
\begin{equation}
\abold = \abold_{\prob} \flatbold^{*} = \flatbold^{*}  \abold_{\prob}.
\end{equation}
We furthermore define a natural transformation $\zetabold : \Ibold \to \abold$ by the diagram
\begin{equation}
\xymatrix{
\Ibold \ar [rrdd] ^{\zetabold}  \ar [rr] ^{\zetabold^{\flat}} \ar [dd] _{\zetabold^{\prob}} &&  \flatbold^{*}  \ar [dd] ^{\flatbold^{*}\zetabold^{\prob}} \\
&& \\
\abold_{\prob} \ar [rr] _{\abold_{\prob}\zetabold^{\flat}} && \;\abold    \\
}
\label{eq:definition of boldzeta}
\end{equation}
where the outer square commutes.
For each $\Qbold \in \Cboldhat$,
the morphism $\Qbold \xrightarrow{\zetabold_{\Qbold}} \abold \Qbold$ is explicitly given as,
\begin{equation}
(\zetabold_{\Qbold})_{(V,r)}= \Hom_{\Cboldhat}(\unitbold_{\downarrow (\flat(V),\mathring{r})} \rightarrowtail  \unitbold_{\downarrow (V,r )},\Qbold).
\end{equation} 

Conditions for a presheaf $\Qbold \in \Cbold$ to be a $\jbold$-sheaf are summarized as follows.
\begin{prp}
The following conditions are equivalent.
\begin{description}
\item[(i)] A presheaf $\Qbold \in \Cboldhat$ is a $\jbold$-sheaf.
\item[(ii)] The morphism $\Qbold \xrightarrow{\zetabold_{\Qbold}} \abold \Qbold$ is isomorphic.
\item[(iii)] For every $(V,r) \in \Cbold$, the presheaves $\Qbold(-,r) \in \Vboldhat$ and $\Qbold(V,-) \in \Ihat$ are a $j$-sheaf and a $j_{\prob}$-sheaf, respectively.
\end{description}
\end{prp}
Proof.
First note that 
$\Qbold \in \Cboldhat$ is a  $\Jbold$-sheaf (hence, a $\jbold$-sheaf)
if and only if
\begin{equation}
\Hom_{\Cboldhat}(\unitbold_{\omegabold}\rightarrowtail \unitbold_{\downarrow(V,r)}, \Qbold):
\Hom_{\Cboldhat}(\unitbold_{\downarrow(V,r)}, \Qbold) \to
\Hom_{\Cboldhat}(\unitbold_{\omegabold}, \Qbold)
\end{equation}
is a bijection for every $(V,r) \in \Cbold$ and $\omegabold \in \Jbold(V,r)$.
Here,
$\unitbold_{\omegabold} \in \Cboldhat$ is defined by
\begin{equation}
\unitbold_{\omegabold}(V',r') :=
\begin{cases}
\{ * \} & (V',r') \in \omegabold, \\
\emptyset & \mbox{otherwise}.\\
\end{cases}
\end{equation}
The condition (ii) readily follows from (i), if we take $\omegabold = \downarrow (\flat(V), \mathring{r}) \in \Jbold(V,r)$. 

Next, we show that (ii) implies (i).
Condition (ii) means that for each $(V,r) \in \Cbold$,
the function $\Hom_{\Cboldhat}(\unitbold_{\downarrow (\flat(V),\mathring{r})}\rightarrowtail \unitbold_{\downarrow (\flat(V),r)}, \Qbold)$ is a bijection.
Therefore, we have the following commutative diagram
\begin{equation}
\xymatrix{
\Hom_{\Cboldhat}(\unitbold_{\downarrow(V,r)}, \Qbold) \ar [rr] ^{(\zetabold_{\Qbold})_{(V,r)}} _{\sim}\ar [rrdd] _{\Qbold(\flat(V)\hookrightarrow V,r) \quad}&&
\Hom_{\Cboldhat}(\unitbold_{\downarrow (\flat(V),\mathring{r})}, \Qbold) \\
&&\\
&& \Hom_{\Cboldhat}(\unitbold_{\downarrow (\flat(V),r)}, \Qbold) \ar [uu] _ {(\zetabold_{\Qbold})_{(\flat(V),r)}} ^{\vertsim}\\
}
\label{eq:boldzetaflatQ}
\end{equation}
Accordingly, the function $(\zetabold^{\flat}_{\Qbold})_{(V,r)}=\Qbold(\flat(V) \hookrightarrow V,r)$ is a bijection,
which means that the presheaf $\Qbold(-,r) \in \Vboldhat$ is a $j$-sheaf.

To verify that (ii) implies $\Qbold(V,-)$ is a $j_{\prob}$-sheaf,
we consider the case where $(\flat(V),r) \not\in \omega$ first.
We define a map $h:\Hom_{\Cbold}(\unitbold_{\omegabold}, \Qbold) \to \Hom_{\Cbold}(\unitbold_{\downarrow (V,r)}, \Qbold)$ as follows:
For every $\alphabold \in \Hom_{\Cbold}(\unitbold_{\omegabold}, \Qbold)$,
$h(\alphabold) \in \Hom_{\Cbold}(\unitbold_{\downarrow (V,r)}, \Qbold)$ is defined by
the diagram
\begin{equation}
\xymatrix{
\unitbold_{\downarrow (\flat(V),\mathring{r})} \ar @{>->} ^{\alphabold|_{\downarrow (\flat(V),\mathring{r}) }} [rrd] \ar @{>->} [d]  &&\\
\unitbold_{\omegabold} \ar @{>->} ^{\alphabold} [rr] \ar @{>->} [d]  &&   \Qbold \\
\unitbold_{\downarrow (V,r)} \ar @{>->} _{h(\alphabold)} [rru] && \\
}
\label{eq:two triangles}
\end{equation}
the upper-half and the outer triangles of which commute.
Namely,
\begin{equation}
h(\alphabold) :=(\zetabold_{\Qbold})^{-1}_{(V,r)}(\alphabold|_{\downarrow (\flat(V),\mathring{r}) }).
\end{equation}
To see that $h$ is bijective, it suffices to check commutativity of the lower-half triangle.
To do so, 
we check the following diagram for  $(W,s) \in \omegabold$:
\begin{equation}
\xymatrix{
\unitbold_{\downarrow (\flat(V), \mathring{r})}(\flat(W),s) \ar @{>->} [rr] ^{(\alphabold|_{\downarrow  (\flat(V), \mathring{r})})_{(\flat(W),s)}} \ar  @{=} [ddd] \ar @{=} [rd] && \Qbold(\flat(W),s) \\
&  \unitbold_{\omegabold} (\flat(W),s) \ar @{>->} [ru] _{\alphabold_{ (\flat(W),s)}} \ar @{=} [d]&  \\
&  \unitbold_{\omegabold} (W,s) \ar @{>->} [rd] ^{\alphabold_{ (W,s)}} & \\
\unitbold_{\downarrow (V,r)}(W,s) \ar @{=} [ru] \ar @{>->} [rr] _{(h(\alphabold))_{(W,s)}}&& \Qbold(W,s) \ar [uuu] _{\Qbold(\flat(W)\hookrightarrow W,s)} ^{\vertsim} \\
} \label{eq:lower triangle}
\end{equation}
In this diagram, the outer square, the upper triangle, and the  right-hand side trapezoid are commutative.
Also, as mentioned above, $\Qbold(\flat(W)\hookrightarrow W,s)$ is bijective.
Therefore, the lower triangle, which is the $(W,s)$-component of the lower triangle of (\ref{eq:two triangles}), is commutative.

If $(\flat(V),r) \in \omegabold$,
define $h$ as
\begin{equation}
h(\alpha) := \Qbold(\flat(V) \hookrightarrow V,r)^{-1}(\alphabold|_{\downarrow (\flat(V),r)}),
\end{equation} 
and replace the morphism at the top of (\ref{eq:lower triangle}) by $\alphabold|_{\downarrow (\flat(V),r)}$.
Then a similar proof holds.

Finally, we show that (ii) $\Leftrightarrow$ (iii).
If (iii) holds,
$\zetabold^{\flat}_{\Qbold}$ and $\zetabold^{\prob}_{\Qbold}$ are isomorphic in (\ref{eq:definition of boldzeta}),
hence, so is $(\flatbold^{*}\zetabold^{\prob})_{\Qbold}$.
Thus, $\zetabold$ is isomorphic.
Conversely, if (ii) holds,
$\zetabold^{\flat}_{\Qbold}$ is isomorphic as shown in (\ref{eq:boldzetaflatQ}),
hence, so is $(\abold_{\prob} \zetabold^{\flat})_{\Qbold}$,
which implies $\zetabold_{\Qbold}$ is isomorphic.
\qed

Obviously, for every $\Qbold \in \Cboldhat$,
$\abold \Qbold$ is a $\jbold$-sheaf.
In fact, the functor $\abold:\Cboldhat \to \Sh_{\jbold} \Cboldhat$ is a sheafificaton functor associated with
 $\jbold$,
and furthermore,  $\zetabold$ is the unit of the adjunction $\abold \dashv \boldsymbol{i}$,
where $\boldsymbol{i}:\Sh_{\jbold} \Cboldhat \hookrightarrow \Cboldhat$ is the inclusion.

We describe the power object $\Pmathbb_{\jbold}\Rbold \equiv \Rbold^{\Omegabold_{\jbold}}\in \Sh_{\jbold}\Cboldhat$
 of a $\jbold$-sheaf $\Rbold$.
For $(V,r) \in \Cbold$, $(\Pmathbb_{\jbold}(\abold \Qbold))(V,r) $ is given by
\begin{eqnarray}
(\Pmathbb_{\jbold}(\abold \Qbold))(V,r) 
& = & \Hom_{\Cboldhat}
((\abold \Qbold)_{\downarrow (V,r)}, \Omegabold_{\jbold \,\downarrow (V,r)}) \nonumber\\
& \simeq &
 \Hom_{\Cboldhat}((\abold \Qbold)_{\downarrow (V,r)}, \Omegabold_{\jbold}) \nonumber\\
& \simeq &
\Hom_{\Cboldhat}(\abold (\Qbold_{\downarrow (V,r)}), \Omegabold_{\jbold}) \nonumber\\ 
& \simeq &
\Sub_{\jbold}(\abold (\Qbold_{\downarrow (V,r)})),
\end{eqnarray}
where from the second line to the third,
we used the fact that  $\overline{(\abold \Qbold)_{\downarrow (V,r)}}=\abold (\Qbold_{\downarrow (V,r)})$
and the fact that $\Omegabold_{\jbold}$ is a $\jbold$-sheaf.
For each $(V',r') \hookrightarrow (V,r) \in \Mor(\Cbold)$,
\begin{equation}
(\Pmathbb_{\jbold}(\abold \Qbold))((V',r') \hookrightarrow (V,r)) =
\Hom_{\Cboldhat}(\abold(\Qbold_{\downarrow (V',r')})  \rightarrowtail \abold( \Qbold_{\downarrow (V,r)}), \Omegabold_{\jbold}) ,
\end{equation}
or, as a map from $\Sub_{\jbold}(\abold (\Qbold_{\downarrow (V,r)})$ to $\Sub_{\jbold}(\abold (\Qbold_{\downarrow (V',r')})$,
\begin{equation}
(\Pmathbb_{\jbold}(\abold \Qbold))((V',r') \hookrightarrow (V,r)) (\Sbold) = \abold (\Sbold_{\downarrow (V',r')}).
\end{equation}

There exists a bijection between $\Sub_{\jbold}(\abold \Qbold)$ and $\Gamma (\Pmathbb_{\jbold}(\abold \Qbold)) $; that is,  every sub-sheaf $\Sbold$ of $\abold \Qbold$ has its name $\unitbold \xrightarrow{\lceil \Sbold \rceil} \Pmathbb_{\jbold}(\abold \Qbold)$
uniquely given by 
\begin{equation}
\lceil \Sbold \rceil_{(V,r)} := \abold (\Sbold_{\downarrow (V,r)}).
\end{equation}

Let $\pibold_{1}:\Cbold \to \Vbold$ be the projection functor with respect to the first argument: 
\begin{equation}
\pibold_{1} (V,r) = V \quad \mbox{and} \quad \pibold_{1}((V',r') \hookrightarrow (V,r)) = V' \hookrightarrow V.
\end{equation}
Then, the pullback functor $\pibold_{1}^{*}:\Vboldhat \to \Cboldhat$ is defined by
\begin{equation}
(\pibold_{1}^{*} Q)(V,r) = Q(V) ,
\end{equation}
\begin{equation}
(\pibold_{1}^{*} Q)((V',r') \hookrightarrow (V,r)) := Q(V' \hookrightarrow V) ,
\end{equation}
and for each $f \in \Mor(\Vboldhat)$,
\begin{equation}
(\pibold_{1}^{*}f)_{(V,r)} = f_{V}.
\end{equation}
Since $\abold_{\prob}\pibold_{1}^{*} Q = \pibold_{1}^{*} Q$ for every $Q \in \Vbold$,
$\pibold_{1}^{*} Q$ is always a $j_{\prob}$-sheaf.
Therefore, $\flatbold^{*} \pibold_{1}^{*} Q = \pibold_{1}^{*} \flat^{*} Q$ is a $\jbold$-sheaf.
In particular, if $Q$ itself is a $j$-sheaf, $\pibold_{1}^{*} Q$ is a $\jbold$-sheaf.

\begin{prp} 
\label{prp:sub- injection}
For every presheaf $Q \in \Vboldhat$ and every $(V,r) \in \Cbold$,
the map defined by 
\begin{equation}
\Sub_{j}(\flat^{*}(Q_{\downarrow V})) \to \Sub_{\jbold}(\abold((\pibold_{1}^{*}Q)_{\downarrow (V,r)}));
S= \flat^{*}(S_{\downarrow V}) \mapsto \abold((\pibold_{1}^{*}S)_{\downarrow (V,r)})
\label{eq:sub- injection}
\end{equation}
is injective.
\end{prp}
Proof.
Note that for each $S \in \Sub_{j}(\flat^{*}(Q_{\downarrow V})) $
and for each $(V',r') \in \Cbold$,
\begin{eqnarray}
\abold((\pibold_{1}^{*}S)_{\downarrow (V,r)}) (V',r') 
& \simeq &
\Hom_{\Cboldhat}(\unitbold_{\downarrow (\flat(V'),\mathring{r'})}, (\pibold_{1}^{*} S)_{\downarrow (V,r)}) \nonumber\\
& = &
\begin{cases}
\Hom_{\Cboldhat}(\unitbold_{\downarrow (\flat(V'),\mathring{r'})}, \pibold_{1}^{*} S)  & (\flat(V'),r') \le_{\Cbold} (V,r) \\
\emptyset & \mbox{otherwise} \\
\end{cases} \nonumber\\
& \simeq &
\begin{cases}
\Hom_{\Vboldhat}(1_{\downarrow \flat(V')}, S)  & (\flat(V'),r') \le_{\Cbold} (V,r) \\
\emptyset & \mbox{otherwise} \\
\end{cases} \nonumber\\
& \simeq &
\begin{cases}
S(\flat(V')) &  (\flat(V'),r') \le_{\Cbold} (V,r) \\
\emptyset & \mbox{otherwise}. \\
\end{cases}
\end{eqnarray}
Suppose that sub-sheaves $S$ and $S'$ of $\flat^{*} (Q_{\downarrow V})$ satisfy
\begin{equation}
\abold((\pibold_{1}^{*}S)_{\downarrow (V,r)}) \simeq \abold((\pibold_{1}^{*}S')_{\downarrow (V,r)}).
\end{equation}
Then, we have, for all $V' \in \Vbold$ such that $\flat(V') \subseteqVbold V$,
\begin{equation}
S(\flat(V')) \simeq  S'(\flat(V')),
\end{equation}
which implies that $S \simeq S'$. 
Thus, they are the same in $\Sub_{j}(\flat^{*}(Q_{\downarrow V})) $.
\qed

The injection (\ref{eq:sub- injection}) gives an inclusion
\begin{equation}
\pibold_{1}^{*}(\Pmathbb_{j}(\flat^{*} Q))(V,r)
=
\Pmathbb_{j}(\flat^{*} Q)(V) \rightarrowtail \Pmathbb_{\jbold}(\abold(\pibold_{1}^{*}Q))(V,r),
\end{equation}
hence, a monic in $\Cboldhat$,
\begin{equation}
\pibold_{1}^{*}(\Pmathbb_{j}(\flat^{*} Q)) \rightarrowtail \Pmathbb_{\jbold}(\abold(\pibold_{1}^{*} Q)).
\label{eq:monic power obj}
\end{equation}
Correspondingly, we can detach the sub-scripts $\downarrow V$ and $\downarrow (V,r)$ from (\ref{eq:sub- injection}), hence, obtain an injection
\begin{equation}
\Sub_{j}(\flat^{*}Q) \to \Sub_{\jbold}(\abold(\pibold_{1}^{*}Q));
S= \flat^{*}S \mapsto \abold(\pibold_{1}^{*}S) \simeq \pibold_{1}^{*}S.
\label{eq:sub-injection full}
\end{equation}

Let $\Sigmabold := \pibold_{1}^{*} \Sigma$.
Then, $\flatbold^{*} \Sigmabold = \pibold_{1}^{*}\flat^{*}\Sigma$ is a $\jbold$-sheaf,
which we call the spectral $\jbold$-sheaf.
It is obvious from (\ref{eq:sub-injection full}) that $\pibold_{1}^{*}$ gives an injection
from $\Sub_{j\,\cl}(\flat^{*}\Sigma)$ to $\Sub_{\jbold \, \cl} (\flatbold^{*} \Sigmabold)$
by $S \mapsto \pibold_{1}^{*} S$,
where $\Sub_{\jbold \, \cl} (\flatbold^{*} \Sigmabold)$ denotes the collection of clopen sub-sheaves of $\flatbold^{*} \Sigmabold$.
Since $\abold \Sigmabold = \flatbold^{*} \Sigmabold$ holds,
$\Sub_{\jbold \, \cl}(\flatbold^{*} \Sigmabold)$ has its internal expression $\Pmathbb_{\jbold \,\cl}(\flatbold^{*} \Sigmabold)$
that is given by 
\begin{equation}
(\Pmathbb_{\jbold \,\cl}(\flatbold^{*} \Sigmabold))(V,r)
=
\Sub_{\jbold \, \cl}(\abold (\Sigmabold_{\downarrow (V,r)}) ).
\end{equation}

From (\ref{eq:monic power obj}) and (\ref{eq:sub-injection full}), 
we have an injection $\Sub_{j \, \cl}(\flat^{*}(\Sigma)) \hookrightarrow \Sub_{\jbold \, \cl}(\flatbold^{*}\Sigmabold))$ and a monomorphism $\pibold_{1}^{*}(\Pmathbb_{j\,\cl}(\flat^{*} \Sigma)) \rightarrowtail \Pmathbb_{\jbold\, \cl}(\abold(\pibold_{1}^{*} \Sigmabold))$
by which physical propositions in $\Sh_{j}\Vboldhat$ are faithfully represented in $\Sh_{\jbold} \Cboldhat$.

Let $\rho$ be a density matrix.
We define $\Tbold^{\rho}_{\jbold} \in \Cboldhat$  by
 \begin{eqnarray}
\Tbold^{\rho}_{\jbold}(V,r) &:=&
\{
\Sbold \in \Sub_{\jbold \, \cl}(\abold(\Sigmabold_{\downarrow (V,r)})) \;|\;
\exists S \in \Tmathbbrhor_{j}(V) \; 
(\pibold_{1}^{*}S \subseteq \Sbold) \}
\nonumber\\
& = &
\{\Sbold \in \Sub_{\jbold \, \cl}(\abold(\Sigmabold_{\downarrow (V,r)})) \;|\;
\Sbold(-,r) \in  \Tmathbb^{\rho,\, r}_{j}(V) \} \nonumber\\ 
& = &
\{
\Sbold \in \Sub_{\jbold \, \cl}(\abold(\Sigmabold_{\downarrow (V,r)})) \;|\;
\tr(\rho \Phat_{\Sbold(V,r)}) \ge r
\}.
\end{eqnarray}
\begin{prp}
The presheaf $\Tbold^{\rho}_{\jbold}$ is  a sub-sheaf of $\Pmathbb_{\jbold \, \cl} (\flatbold^{*}\Sigmabold)$.
\end{prp} 
Proof.
First, we show that  $\Tbold^{\rho}_{\jbold}$ is a sub-presheaf of $\Pmathbb_{\jbold \, \cl} (\flatbold^{*}\Sigmabold)$;
that is, if $\Sbold \in \Tmathbb^{\rho}_{\jbold}(V,r)$ and $(V',r') \le_{\Cbold} (V,r)$,
then $\abold (\Sbold_{\downarrow (V',r')}) \in \Tmathbb^{\rho}_{\jbold}(V',r')$.
To do so, note that we have
\begin{equation}
\abold (\Sbold_{\downarrow (V',r')}) (V',r') = \Sbold(V',r') \supseteq \Sbold(V',r).
\end{equation}
Therefore,
\begin{equation}
\tr(\rho \Phat_{\abold (\Sbold_{\downarrow (V',r')}) (V',r')})=
\tr(\rho \Phat_{\Sbold(V',r')}) 
 \ge 
\tr(\rho \Phat_{\Sbold(V',r)})  \ge 
\tr(\rho \Phat_{\Sbold(V,r)}) 
 \ge 
r 
 \ge 
r'.
\end{equation}

Obviously, $\Tmathbb^{\rho}_{\jbold}(-,r) \in \Vboldhat$ is a $j$-sheaf.
In fact, $\zetabold^{\flat}_{\Tmathbb^{\rho}_{\jbold}}$ has the inverse morphism,
which assigns to each $\Sbold\in   (\flatbold^{*} \Tmathbb^{\rho}_{\jbold})(V,r) $
a $\jbold$-sheaf $\Sbold' \in \Tmathbb_{\jbold}^{\rho}(V,r)$ defined by
\begin{equation}
\Sbold'(V',r') := 
\begin{cases}
\Sbold(\flat(V'),r')  & (\flat(V'),r') \le_{\Cbold} (V,r) \\
\emptyset & \mbox{otherwise}
\end{cases}
\end{equation} 
and
\begin{equation}
\Sbold'((V'',r'') \hookrightarrow (V',r')) := \Sbold((\flat(V''),r'') \hookrightarrow (\flat(V'),r')).
\end{equation}

We show that $\Tmathbb^{\rho}_{\jbold}(V,-)$ is a $j_{\prob}$-sheaf.
To do so, note that the diagram
\begin{equation}
\xymatrix{
\Hom_{\Cboldhat}(\unitbold_{\downarrow(V,r)}, \Tmathbb^{\rho}_{\jbold}) 
\ar [rr] ^{\zetabold^{\prob}_{\Tmathbb^{\rho}_{\jbold}}} \ar [dd] _{\vertsim} && 
\Hom_{\Cboldhat}(\unitbold_{\downarrow(V, \mathring{r})},\Tmathbb^{\rho}_{\jbold}) 
\ar [dd] ^{\vertsim} \\
&& \\
\Hom(1_{\downarrow r}, \Tmathbb^{\rho}_{\jbold}(V,-)) \ar [rr] ^{(\zeta^{\prob}_{\Tmathbb^{\rho}_{\jbold}(V,-)})_{r} }
\ar [dd] _{\vertsim}
&& \Hom(1_{\downarrow \mathring{r}}, \Tmathbb^{\rho}_{\jbold}(V,-)) \\
&&\\
\Tmathbb^{\rho}_{\jbold}(V,r) \ar [rruu]&& \\
}
\end{equation}
defines a map 
\begin{equation}
\Tmathbb^{\rho}_{\jbold}(V,r) \to \Hom(1_{\downarrow \mathring{r}}, \Tmathbb^{\rho}_{\jbold}(V,-));
\Sbold \mapsto (\abold(\Sbold_{\downarrow (V,s)}))_{s \in \downarrow \mathring{r}} \, .
\label{eq:zetaT-j-rho}
\end{equation}
On the other hand, we can define a map from $\Hom(1_{\downarrow \mathring{r}}, \Tmathbb^{\rho}_{\jbold}(V,-))$
to $\Tmathbb^{\rho}_{\jbold}(V,r)$ as follows.
Let us take an arbitrary $(\Sbold_{r'})_{r' \in \downarrow \mathring{r}} \in \Hom(1_{\downarrow \mathring{r}},\Tmathbb^{\rho}_{\jbold}(V,-))$.
For every $V' \subseteqVbold V$ and $s' \le s \le r'$,
we have
\begin{equation}
\Sbold_{s'}(V',s') = \abold((\Sbold_{s})_{\downarrow (V',s')}) (V',s') 
 = \Sbold_{s}(\flat(V'),s') = \Sbold_{s}(V',s') \supseteq \Sbold_{s}(V',s).
\end{equation}
So we define a $\jbold$-sheaf $\Sbold$ by
\begin{equation}
\Sbold(V',s') := 
\begin{cases}
\displaystyle{\bigcap_{s \in \downarrow \mathring{r'}} }\{ \Sbold_{s}(V',s) \} & (\flat(V'),r') \le_{\Cbold} (V,r) \\
& \\
\emptyset & \mbox{otherwise} \,.\\
\end{cases}
\end{equation}
In fact, it is not difficult to see that $\Sbold \in \Sub_{\jbold \, \cl}(\abold(\Sigmabold_{\downarrow (V,r)}))$.

Such a $\jbold$-sheaf $\Sbold$ satisfies,  for all $r' \in \downarrow \mathring{r}$,
\begin{equation}
\Sbold(V,r) = \bigcap_{s \in \downarrow \mathring{r}}\{ \Sbold_{s}(V,s) \} 
\subseteq \bigcap_{s \in \downarrow r'}\{ \Sbold_{s}(V,s) \} = \Sbold_{r'}(V,r'),
\end{equation}
hence,
\begin{equation}
\Phat_{\Sbold(V,r) } = \bigwedge_{s \in \downarrow \mathring{r}}\{\Phat_{\Sbold_{s}(V,s)} \}
\preceq \bigwedge_{s \in \downarrow r'}\{\Phat_{\Sbold_{s}(V,s)} \} = \Phat_{\Sbold_{r'}(V,r')}.
\end{equation}
Since $\Sbold_{r'} \in \Tmathbb^{\rho}_{\jbold}(V,r')$,
we have $\tr(\rho \Phat_{\Sbold_{r'}(V,r')}) \ge r' $.
Thus, 
\begin{equation}
\tr(\rho \Phat_{\Sbold(V,r) }) =
\inf_{s \in \downarrow \mathring{r}} \tr (\rho \Phat_{\Sbold_{s}(V,s) })
= 
\lim_{r' \uparrow r} \tr (\rho \Phat_{\Sbold_{r'}(V,r')}) \ge \lim_{r' \uparrow r} r'
 =r,
\end{equation}
which implies that $\Sbold \in \Tmathbb^{\rho}_{\jbold}(V,r)$.
Obviously the above mentioned correspondence
$(\Sbold_{r'})_{r' \in \downarrow \mathring{r}} \mapsto \Sbold$
is the inverse of (\ref{eq:zetaT-j-rho}),
from which the inverse morphism of $\zetabold_{\Tmathbb^{\rho}_{\jbold}}$ is obtained.
\qed 

Note that $\Tmathbb^{\rho}_{\jbold}$ is  a truth sheaf,
since $\Tmathbb^{\rho}_{\jbold}(V,r)$ is a filter for any $(V,r) \in \Cbold$.
In fact, it is not difficult to see that
$\Tmathbb^{\rho}_{\jbold}(V,r)$ is the least filter including $\Tmathbb_{j}^{\rho,\, r}(V)$.
So, we can adopt $\Tmathbb^{\rho}_{\jbold}$ as the truth sheaf associated with $\rho$.

Let $\Sbold \in \Sub_{\jbold \, \cl} (\flatbold^{*} \Sigmabold)$ be a physical proposition.
The truth-value $\nubold_{\jbold}^{\rho}(\Sbold)$ of $\Sbold$
under the truth-object $\Tmathbb^{\rho}_{\jbold}$ is given as
\begin{eqnarray}
(\nubold^{\rho}_{\jbold}(\Sbold))_{(V,r)}
& = &
\{
(V',r') \le_{\Cbold} (V,r) \;|\;
\lceil \Sbold \rceil_{(V',r')} \in \Tbold^{\rho}_{\jbold}(V',r')
\} \nonumber\\
& = &
\{
(V',r') \le_{\Cbold} (V,r) \;|\;
\tr (\rho \Phat_{\Sbold(V',r')} ) \ge r'
\}.
\end{eqnarray}

D\"{o}ring and Isham \cite{DI12} gave topos theoretic formulation of classical probabilities.
And further, they extended it to the case of quantum probabilities.
We can give the sheaf-based counterpart by the following commutative diagram:
\begin{equation}
\xymatrix{
\Sub_{j \,\cl}(\flat^{*}\Sigma) \ar  [rr] ^{\mu_{j}^{\rho}} \ar [dd] _{\pibold^{*}_1} && \Gamma [0,1]^{\succeq}_{j} \ar [dd] ^{\ell'} \\
&& \\
\Sub_{\jbold \,\cl}(\flatbold^{*} \Sigmabold) \ar [rr] _{\nubold_{\jbold}^{\rho}} && \Gamma \Omegabold_{\jbold} \\
}
\label{eq:key diagram} 
\end{equation}
Here, we define the map $\ell'$:
For each $\gamma \in\Gamma [0,1]^{\succeq}_{j} $,
$\ell'(\gamma) \in \Gamma \Omegabold_{\jbold} $ is defined by
\begin{equation}
(\ell'(\gamma))_{(V,r)} := \{(V',r') \le_{\Cbold} (V,r) \;|\; h_{\gamma}(\flat(V')) \ge r'\}.
\end{equation}
Commutativity of diagram (\ref{eq:key diagram}) can be shown by direct calculations.
When $\gamma = \mu^{\rho}_{j}(S)$ ($S \in \Sub_{j \cl}(\flat^{*}\Sigma)$),
we have
\begin{equation}
h_{\gamma}(\flat(V')) =\tr(\rho \Phat_{S(\flat(V'))}) = \tr(\rho \Phat_{S(V')}).
\end{equation}
Therefore, 
\begin{equation}
(\ell'(\mu_{j}^\rho(S)))_{(V,r)}
=
\{
(V',r') \le_{\Cbold} (V,r) \;|\;
\tr(\rho \Phat_{S(V')}) \ge r'
\}.
\end{equation}
On the other hand,
\begin{eqnarray}
(\nubold_{\jbold}^{\rho}(\pibold_{1}^{*} S))_{(V,r)}
&= &
\{
(V',r') \le_{\Cbold} (V,r) \;|\;
\abold((\pibold_{1}^{*}S)_{\downarrow (V',r')}) (-,r') \in \Tmathbb_{j}^{\rho,\,r'}(V')
\} \nonumber\\
&=&
\{
(V',r') \le_{\Cbold} (V,r) \;|\; \tr(\rho \Phat_{\pibold_{1}^{*}S(V',r')}) \ge r' \}\nonumber\\
& = &
\{
(V',r') \le_{\Cbold} (V,r) \;|\; \tr(\rho \Phat_{S(V')}) \ge r' 
\} .\nonumber\\
\end{eqnarray}
Thus, the diagram (\ref{eq:key diagram}) is commutative.

\def\flatVboldhat{\widehat{\flat(\Vbold)}}

\section{Reduced quntum theory on presheaves on selected contexts}
\label{presheaf-based reduced theory}

\def\flattilde{\tilde{\flat}}
Let us recall section 2.
We gave reduced topos quantum theory based on sheaves induced by 
an context selector
and showed that the reduced theory well works for $\flat$-selected operators.
Also, structural relationship to the original presheaf theory of D\"{o}ring and Isham has been clarified.
In this section, we further procceed to formulate a more manageable framework on which 
topos quantum theory equivalent to the sheaf-based one can be constructed.
In fact, though $j$-sheaves are defined on all over $\Vbold$,
they can be determined by their values on $\flat(\Vbold)$,
which is a full and faithful sub-category of $\Vbold$.
The topos quantum theory, therefore, can be formulated on the reduced presheaf topos 
$\flatVboldhat \equiv \Set^{\flat(\Vbold)^{\op}}$ on $\flat(\Vbold)$.

%
We write $\flattilde$ for the functor $\flat$ regarded as $\flat:\Vbold \to \flat(\Vbold)$. 
Then, the pullback functor  $\flattilde^{*}: \flatVboldhat \to \Vboldhat$
is defined by
\begin{equation}
(\flattilde^{*}X)(V) := X(\flat(V)),
\quad
(\flattilde^{*}X)(V' \hookrightarrow V) := X(\flat(V') \hookrightarrow \flat(V)),
\end{equation}
and for each $X \xrightarrow{k} Y \in \Mor (\flatVboldhat)$,
\begin{equation}
(\flattilde^{*}k)_{V} := k_{\flat(V)}.
\end{equation}
Note that $\flattilde^{*}$ is full and faithful  as a functor from $\flatVboldhat$ to $\Sh_{j}\Vbold$.
Furthermore,  $\flattilde^{*}$ is left-adjoint to the functor $\flattilde_{*}:\Vboldhat \to\flatVboldhat$ 
defined by
\begin{equation}
\flattilde_{*}Q := Q|_{\flat(\Vbold)}
\quad \mbox{and} \quad 
\flattilde_{*}(Q \xrightarrow{f} R) := Q|_{\flat(\Vbold)} \xrightarrow{f|_{\flat(\Vbold)}} R|_{\flat(\Vbold)} ,
\end{equation}
and the associated unit is the identity $I_{\flatVboldhat} \to I_{\flatVboldhat}$.
\begin{prp}
Let $\Omega_{\flat}$ be the sub-object classifier of $\flatVboldhat$.
Then it follows that 
\begin{equation}
\flattilde_{*} \Omega_{j} \simeq \Omega_{\flat}.
\end{equation}
\end{prp}
Proof.
The morphism $\flattilde_{*} \Omega_{j} \xrightarrow{\xi} \Omega_{\flat}$ defined by
\begin{equation}
\xi_{\flat(V)}:(\flattilde_{*} \Omega_{j})(\flat(V)) = \Omega_{j}(\flat(V)) \to \Omega_{\flat}(\flat(V)); \omega \mapsto \omega \cap \flat(\Vbold)
\end{equation}
 has the inverse  $\Omega_{\flat} \to \flattilde_{*} \Omega_{j}$ whose $\flat(V)$-components are given by
\begin{equation}
\Omega_{\flat}(\flat(V)) \to (\flattilde_{*} \Omega_{j})(\flat(V)): \omega_{\flat} \mapsto \{ V' \subseteqVbold \flat(V) \;|\; \flat(V') \in \omega_{\flat} \}.
\end{equation}
\qed

Let $1_{\flat}$ be the terminal object of $\flatVboldhat$.
Then, we have $\flattilde^{*} 1_{\flat}= 1$.
Sheaf-based truth-values can be translated to presheaf-based ones in $\flatVboldhat$ as
\begin{eqnarray}
\Gamma \Omega_{j} &\equiv&\Hom(1, \Omega_{j})  \nonumber\\
& \simeq &
\Hom (\flattilde^{*}1_{\flat},  \Omega_{j} ) \nonumber\\
 & \simeq &
 \Hom_{\flatVboldhat} (1_{\flat}, \flattilde_{*}\Omega_{j} ) \nonumber\\
& \simeq &
 \Hom_{\flatVboldhat} (1_{\flat}, \Omega_{\flat}) \equiv \Gamma \Omega_{\flat} .
\end{eqnarray}
In the following, we complete the translation rule between the quantum theories on $\Sh_{j} \Vboldhat$ and $\flatVboldhat$.

For each $X\in \flatVboldhat$,  the collection $\Sub_{\,\flat}(X)$ of its sub-objects is 
internalized to $\flatVboldhat$ by
$\Pmathbb_{\flat} X \equiv X ^{\Omega_{\flat}}$,
which is expressed as
\begin{equation}
(\Pmathbb_{\flat} X)(\flat(V))
=
\Sub_{\,\flat}(X_{\Downarrow \flat(V)})
\end{equation}
and
\begin{equation}
(\Pmathbb_{\flat} X)(\flat(V') \hookrightarrow \flat(V)):
\Sub_{\,\flat}(X_{\Downarrow \flat(V')}) \to \Sub_{\,\flat}(X_{\Downarrow \flat(V)});
Z \mapsto Z_{\Downarrow \flat(V')},
\end{equation}
where $X_{\Downarrow \flat(V)}$ means the restriction of $X$ to $\flat(\Vbold) \, \cap \downarrow \flat(V)$.
\begin{prp}
For each presheaf $Q \in \Vboldhat$,
we have
\begin{equation}
\Pmathbb_{\,\flat} (\flattilde_{*} Q)   \simeq    \flattilde_{*}(\Pmathbb_{j} (\flat^{*}Q)) 
.
\end{equation}
\end{prp}
Proof.
The two maps
\begin{equation}
(\Pmathbb_{\, \flat}(\flattilde_{*} Q)) (\flat(V)) \to   \flattilde_{*}(\Pmathbb_{j} (\flat^{*}Q)) (\flat(V)) ;
Z \mapsto \flattilde^{*} Z
\label{eq:translation of proposition}
\end{equation}
and
\begin{equation}
\flattilde_{*}(\Pmathbb_{j} (\flat^{*}Q)) (\flat(V)) \to (\Pmathbb_{\, \flat}(\flattilde_{*} Q)) (\flat(V)) ;
Z \mapsto \flattilde_{*} Z 
\end{equation}
 are inverse each other.
\qed
In particular, it is easy to see that the map (\ref{eq:translation of proposition}) to the presheaf $\Pmathbb_{\flat\,\cl}(\flattilde_{*} \Sigma)$,  an internalization of the collection $\Sub_{\flat \, \cl} (\flat^{*}\Sigma)$ 
of clopen sub-objects of $\flattilde_{*}\Sigma$,
gives an isomorphism
\begin{equation}
\Pmathbb_{\flat \, \cl} (\flattilde_{*} \Sigma) \xrightarrow{\sim} \flattilde_{*}(\Pmathbb_{j \, \cl} (\flat^{*}\Sigma)),
\label{eq:vartheta}
\end{equation}
which we denote by $\vartheta_{\Sigma}$.

The translation $\Tmathbb_{\flat}$ of a truth sheaf $\Tmathbb_{j}$ is defined as the pullback
$\vartheta^{-1}(\flattilde_{*} \Tmathbb_{j})$ of $\Tmathbb_{j}$ along the morphism  $\Pmathbb_{\flat \, \cl} (\flattilde_{*} \Sigma) \xrightarrow{\vartheta} \flattilde_{*}(\Pmathbb_{j \, \cl} (\flat^{*}\Sigma))$.
That is,
\begin{eqnarray}
\Tmathbb_{\flat}(\flat(V)) & :=&
(\vartheta^{-1}(\flattilde_{*} \Tmathbb_{j}))(\flat(V)) \nonumber\\
& = & 
\{
Z \in \Pmathbb_{\flat \, \cl}((\flattilde_{*} \Sigma)_{\Downarrow \flat(V)}) \;|\;
\flattilde^{*} Z \in \Tmathbb_{j}(\flat(V))
\}.
\end{eqnarray}
The following proposition can be easily shown.
\begin{prp}
Let $\tau_{\flat}$ be the characteristic morphism of $\Tmathbb_{\flat} \rightarrowtail \Pmathbb_{\flat\,\cl}(\flattilde_{*} \Sigma)$.
Then, the diagram
\begin{equation}
\xymatrix{
\Pmathbb_{\flat \, \cl}(\flattilde_{*} \Sigma) \ar [rr] ^{\tau_{\flat}} \ar [dd] ^{\vertsim} _{\vartheta_{\Sigma}}&& \Omega_{\flat} \\
&& \\
\flattilde_{*} \Pmathbb_{j \, \cl} (\flat^{*} \Sigma) \ar [rr] _{\flattilde_{*} \tau_{j}} && \flattilde_{*} \Omega_{j} \ar [uu] _{\xi} ^{\vertsim}
}
\end{equation}
commutes.
\end{prp}

Thus, we are led to the following commutative diagram in $\flatVboldhat$:
\begin{equation}
\xymatrix{
&& & \flattilde_{*}\Tmathbb_{j}  \ar @{>->} [ld] \ar [rr] ^{!} \ar @{>.>} [dd] ^(0.35){\vertsim }&& 1_{\flat} \ar @{>->} [ld] _{\flattilde_{*} \true_{j}} \ar @{=} [dd] \\
&& \flattilde_{*}\Pmathbb_{j \, \cl} (\flat^{*}\Sigma) \ar @{>->} [dd] _{\vartheta^{-1}_{\Sigma}} ^{\vertsim} \ar [rr] _(0.4){\flattilde_{*} \tau_{j}}&& \flattilde_{*} \Omega_{j} \ar @{>->} [dd] ^(0.35){\xi} _(0.35){\vertsim} & \\
1_{\flat}  \ar @{>->} [urr] ^{\flattilde_{*} \lceil P_{j} \rceil_{j}}  \ar @{>->} [drr] _{\lceil P_{\flat} \rceil_{\flat}}&&& \Tmathbb_{\flat} \ar @{.>} [rr] _(0.4){!} \ar @{>.>} [ld] && 1_{\flat}  \ar @{>->} [ld] ^{\true_{\flat}} \\
&& \Pmathbb_{\flat\, \cl}(\flattilde_{*} \Sigma) \ar [rr] _{\tau_{\flat}} && \Omega_{\flat} &  \\
}
\end{equation}
Since the top square is the image of  pullback diagram in (\ref{eq:pullback j}) by the right-adjoint functor $\flattilde_{*}$,
it is a pullback, and hence, so is the bottom square.
Consequently, the latter assigns to each physical proposition represented by
a global element of $\Pmathbb_{\flat \, \cl}(\flattilde_{*} \Sigma)$ a global element of $\Omega_{\flat}$ as a truth-value under the truth presheaf $\Tmathbb_{\flat}$.

Every physical proposition $P_{j} \in \Sub_{j \, \cl}(\flat^{*} \Sigma)$ 
and its translation $P_{\flat} \in \Sub_{\flat \, \cl}(\flattilde_{*} \Sigma)$ is related each other as
\begin{equation}
\lceil P_{\flat} \rceil_{\flat} = \vartheta_{\Sigma}^{-1} \circ \flattilde_{*} \lceil P_{j} \rceil_{j} ,
\end{equation} 
and their truth-values $\nu_{j}(P_{j};\Tmathbb_{j}) \in \Gamma \Omega_{j}$ and $\nu_{\flat}(P_{\flat};\Tmathbb_{\flat})\in \Gamma_{\flat} \Omega_{\flat}$ are related as
\begin{equation}
\nu_{\flat}(P_{\flat};\Tmathbb_{\flat}) = \xi \circ \flattilde_{*} ( \nu_{j}(P_{j};\Tmathbb_{j})).
\end{equation}

Let $\delta_{\flat}$ be the daseinization operator restricted to $\flat(\Vboldhat)$.
That is, for any projection operator $\Phat$, $\delta_{\flat}(\Phat)$ is defined by
\begin{equation}
\delta_{\flat}(\Phat)_{\flat(V)} := \bigwedge \{\alphahat \in \Pcal(\flat(V)) : \alphahat \succeq \Phat\}
\end{equation}
or defined as its equivalent form interpreted as a clopen sub-object of $\flattilde_{*} \Sigma$ by
\begin{equation}
\delta_{\flat}(\Phat)(\flat(V)) := 
\{
\sigma \in (\flattilde_{*}\Sigma)(\flat(V)) \;|\; \sigma (\delta_{\flat}(\Phat)_{\flat(V)}) =1
\}.
\end{equation}
Further, for each density matrix $\rho$ and $r \in [0,1]$,
we define $\Tmathbb_{\flat}^{\rho,\, r} \subseteq \Pmathbb_{\flat\, \cl}(\flattilde_{*} \Sigma)$ by
\begin{equation}
\Tmathbb_{\flat}^{\rho,\, r}(\flat(V))
:= 
\{Z \in \Sub_{\flat \, \cl} (\flattilde_{*} \Sigma)_{\Downarrow \flat(V)} \;|\;
\tr (\rho \Phat_{Z(\flat(V))})  \ge r \}.
\end{equation}
Then, we can easily prove the following propositions.
\begin{prp}
For each projection operator $\Ehat$, $\delta_{\flat}(\Phat)$ is the translation of $\delta_{j}(\Phat)$.
\end{prp}
\begin{prp}
The truth presheaf $\Tmathbb_{\flat}^{\rho, \, r}$ is the translation of $\Tmathbb_{j}^{\rho, \, r}$.
\end{prp}
Thus, we obtain such a topos quantum theory on presheaves on $\flat(\Vbold)$ as is equivalent to sheaf quantum theory on $\Sh_{j}\Vboldhat$.




\appendix

\section{Translation rules between presheaf-based and $j$-sheaf-based topos quantum theories}
\label{sec:translation rule}

Nakayama \cite{N14} gave translation rules of truth-values, propositions, and truth objects
between the presheaf-based topos formulation and the sheaf-based one.
In this appendix, we summarize in terms of the spectral presheaf and 
the spectral sheaf.
(Nakayama \cite{N14} formulated in terms of the outer presheaf and the outer sheaf.)

Physical propositions $S \in \Sub_{\cl} (\Sigma)$ and $S_{j} \in \Sub_{j \,\cl}(\flat^{*}\Sigma)$
are said to be each other's translation if and only if
\begin{equation}
\flat^{*} \Sigma = \Sigma_{j}.
\end{equation}
We define a morphism $\flat^{*}(\Pmathbb_{\cl} \Sigma) \xrightarrow{\varrho_{\Sigma}} \Pmathbb_{j \, \cl}(\flat^{*} \Sigma)$
by 
\begin{equation}
(\varrho_{\Sigma})_{V}:\Sub_{\cl}(\Sigma_{\downarrow \flat(V)}) \to \Sub_{j \, \cl}(\flat^{*}(\Sigma_{\downarrow V})):S \mapsto \flat^{*}S.
\end{equation}
Then, truth objects $\Tmathbb \subseteq \Pmathbb_{\cl} \Sigma$ and $\Tmathbb_{j} \subseteq \Pmathbb_{j\, \cl} (\flat^{*} \Sigma)$ are each other's translation if and only if
\begin{equation}
\flat^{*} \Tmathbb = \varrho^{-1}_{\Sigma}(\Tmathbb_{j}).
\label{eq:translation of truth object}
\end{equation}
Here, $\varrho_{\Sigma}^{-1}(\Tmathbb_{j}) \subseteq \flat^{*} (\Pmathbb_{\cl} \Sigma)$ is
the pullback of $\Tmathbb_{j}$ along the morphism $\varrho_{\Sigma}$ that is defined by
\begin{equation}
\varrho_{\Sigma}^{-1}(\Tmathbb_{j})(V) :=
\{  
S \in \Sub_{\cl} (\Sigma_{\downarrow \flat(V)})
\;|\;
\flat^{*} S \in \Tmathbb_{j}(V)
\}.
\end{equation}
Finally, the translation relation between truth-values $\nu \in \Gamma \Omega$ and $\nu_{j} \in \Gamma \Omega_{j}$ are defined by
\begin{equation}
r  \circ \nu = \nu_{j},
\label{eq:translation of truth-value}
\end{equation}
where $r$ is given by the epi-mono factorization of $j$:
\begin{equation}
\xymatrix{
\Omega \ar [rr] ^{j} \ar @{->>} [rdd] _{r} && \Omega \\
&& \\
& \Omega_{j} \ar @{>->} [ruu]& \\
}
\end{equation}
Nakayam \cite{N14} showed that the above-menstioned translation rules are consistent in the sense that they satisfy
\begin{equation}
r \circ \nu(P;\Tmathbb) = \nu_{j}(P_{j};\Tmathbb_{j}).
\end{equation}

From the sheaf-based viewpoint,
different presheaf-based propositions, truth-objects, and truth-values represent the same ones;
the presheaf based spaces of them are coarse-grained and reduced by the translations.
The degree of the coarse-graining is clarified by Nakayama \cite{N14},
as summarized below.

Every presheaf-based truth-value $\nu$ is translated to a sheaf-based one $r \circ \nu$.
Conversely, for each $\nu_{j}$, a lot of $\nu$'s are translated to $\nu_{j}$ by (\ref{eq:translation of truth-value}).
Let $\gammabold (\nu_{j})$ be the space of such $\nu$'s:
\begin{equation}
\gammabold(\nu_{j}) := \{ \nu \in \Gamma \Omega
\;|\;
r \circ \nu = \nu_{j} \}.
\end{equation}
Then, it can be explicitly given as
\begin{equation}
\gammabold(\nu_{j}) =  \{\nu \in \Gamma \Omega
\;|\;
\gammaboldmin(\nu_{i}) \le \nu \le \gammaboldmax(\nu_{i})
\},
\end{equation}
where $\gammaboldmax(\nu_{j})$ and $\gammaboldmin(\nu_{i}) $ are defined by
\begin{equation}
\gammaboldmax(\nu_{j})  := 
\xymatrix{1 \ar @{>->} [r] ^{\nu_{j}} &  \Omega_{j} \ar @{>->} [r] & \Omega \\},
\end{equation}
and
\begin{equation}
(\gammaboldmin(\nu_{i}))_{V}
:=
\{
V' \subseteqVbold V\;|\;
(\nu_{j})_{V} \cap \Ucalflat(V') \neq \emptyset
\}.
\end{equation}
Here, $\Ucalflat$ is defined as
\begin{equation}
\Ucalflat(V) := \{W \in \Vbold \;|\; V \subseteqVbold \flat(W) \}.
\end{equation}

Each preresheaf proposition $P \in \Sub_{\cl} (\Sigma)$ is translated to a sheaf proposition $\flat^{*} P$.
For each sheaf proposition $P_{j} \in \Sub_{j \, \cl}(\flat^{*}\Sigma)$,
the space $\imathbold(P_{j})$ of presheaf propositions that are translated to $P_{j}$ is given as
\begin{eqnarray}
\imathbold(P_{j}) 
& \equiv &
\{ P \in \Sub_{\cl} (\Sigma) \;|\;
\flat^{*} P = P_{j}\}  \nonumber\\
& = &
\{P \in \Sub_{\cl} (\Sigma) \;|\
\imathboldmin(P_{j}) \subseteq P \subseteq \imathboldmax(P)
\}.
\end{eqnarray}
Here, $\imathboldmax(P)$ and $\imathboldmin(P)$ are defined by
\begin{equation}
\imathboldmax(P)(V) := 
\{
\sigma \in \Sigma(V) \;|\;
\sigma(\Phat_{P_{j}(V)}) = 1
\},
\end{equation}
and 
\begin{equation}
\imathboldmin(P)(V) := 
\begin{cases}
\{
\sigma \in \Sigma(V) \;|\;
\sigma(\bigvee \{ \delta (\Phat_{P_{j}(W)} )_{V} \}_{W \in \Ucalflat(V) }) = 1
\} & \mbox{if } \Ucalflat(V) \neq \emptyset \\
\emptyset  & \mbox{if } \Ucalflat(V) = \emptyset \\
\end{cases}.
\end{equation}

Nakayama \cite{N14} postulated that truth objects should satisfy the filter condition context-wise;
$\Tmathbb \subseteq \Pmathbb_{\cl} \Sigma$ can be a truth presheaf
if $\Tmathbb (V)$ is a filter with respect to its lattice structure given by inclusion. 
All of such  truth presheaves $\Tmathbb \in \Sub_{{\rm filt}}(\Pmathbb_{\cl} \Sigma)$ are not translated to truth sheaves in $\Sh_{j}(\Vboldhat)$.
Let $\jmathbold(\Tmathbb_{j})$ be the family of truth presheaves translated to a truth sheaf $\Tmathbb_{j}$.
Then, we have
\begin{eqnarray}
\jmathbold(\Tmathbb_{j}) 
& := &
\{
\Tmathbb \in \Sub_{{\rm filt}}(\Pmathbb_{\cl} \Sigma) \;|\;
\flat^{*} \Tmathbb  = (\varrho_{\Sigma})^{-1}(\Tmathbb_{j})
\} \nonumber\\
& = &
\{
\Tmathbb \in \Sub_{{\rm filt}}(\Pmathbb_{\cl} \Sigma) \;|\;
\jmathboldmin (\Tmathbb_{j}) \subseteq \Tmathbb \subseteq \jmathboldmax (\Tmathbb_{j}) 
\} .
\end{eqnarray}
Here, $\jmathboldmax (\Tmathbb_{j}) $ is defined by
\begin{equation}
\jmathboldmax (\Tmathbb_{j}) (V) :=
\{
S \in \Sub_{\cl} (\Sigma_{\downarrow V}) \;|\;
\flat^{*}(S_{\downarrow V}) \in \Tmathbb_{j}(V)
\}.
\label{eq:jmathboldmax}
\end{equation}
To give an  expression for $\jmathboldmin (\Tmathbb_{j})$,
we define $\Rmathbb_{V;W}$ ($V \subseteqVbold W$) and $\Rmathbb_{V}$ by
\begin{equation}
\Rmathbb_{V;W} := 
\{
S_{\downarrow V} \in \Sub_{\cl}(\Sigma_{\downarrow V}) \;|\;
S \in ((\varrho_{\Sigma})^{-1}(\Tmathbb_{j})) (W)
\}
\end{equation}
and
\begin{equation}
\Rmathbb_{V} := \bigcup \{ \Rmathbb_{V;W} \}_{W \in \Ucalflat(V)},
\end{equation}
respectively.
Then,  $\jmathboldmin (\Tmathbb_{j})$ is defined as
\begin{equation}
( \jmathboldmin (\Tmathbb_{j}))(V) :=
\begin{cases}
\Fcal_{V}(\Rmathbb_{V}) & \mbox{if } \Ucalflat(V) \ne \emptyset \\
\{ \Sigma_{\downarrow V} \} & \mbox{if } \Ucalflat(V) = \emptyset , \\
\end{cases}
\end{equation}
where $\Fcal_{V}(\Rmathbb_{V})$ is the smallest filter in $\Sub_{\cl}(\Sigma_{\downarrow V})$.



\end{document}